\documentclass{article}

\usepackage{amsmath}
\usepackage{amssymb}
\usepackage{amsthm}
\usepackage{dsfont,amsfonts}
\usepackage{mathrsfs}
\usepackage{epstopdf}

\usepackage{natbib}
\usepackage{graphicx}
\usepackage{url}
\usepackage{Sweave}

\usepackage{abbreviationsOjaNP-ext}

\newcommand{\id}{1\hspace{-2.5mm}{1}}
\newcommand{\pkg}{}
\newcommand{\proglang}{}
\newcommand{\code}[1]{{\small\texttt{#1}}}

\newtheorem{mytheorem}{Theorem}
\newtheorem{mydef}{Definition}

\def\abs{\textrm{abs}}

\providecommand{\bo}{\mathbf}
\providecommand{\bs}{\boldsymbol}

 \author{Daniel Fischer\\
        Natural Resources \\Institute Finland (Luke)\\
        Finland\\
         and\\
         School of Health Sciences\\
         University of Tampere\\
         Finland\\
       \and
         Karl Mosler\\
         Institute of Econometrics\\ and Statistics\\
         University of Cologne\\
         Germany
       \and
         Jyrki M\"ott\"onen\\
         Department of Social Research\\
         University of Helsinki\\
         Finland
       \and
         Klaus Nordhausen\\
         Department of Mathematics\\ and Statistics\\
         University of Turku\\
         Finland\\
         and\\
         School of Health Sciences\\
         University of Tampere\\
        Finland\\
       \and
         Oleksii Pokotylo\\
         Cologne Graduate School\\
         University of Cologne\\
         Germany
       \and
 	    Daniel Vogel\\
 			Institute of Complex Systems\\ and Mathematical Biology\\
 	    University of Aberdeen\\
 	    United Kingdom
 	    }
\title{Computing the Oja Median in \proglang{R}:\\ The Package \pkg{OjaNP} }

\begin{document}

\maketitle

\section*{Abstract}
   The Oja median is one of several extensions of the univariate median to the multivariate case. It has many nice properties,
   but is computationally demanding. In this paper, we first review the properties of the Oja median and compare it to other multivariate medians. Afterwards we discuss four algorithms to compute the Oja median, which are implemented in our
   \proglang{R}-package \pkg{OjaNP}. Besides these algorithms, the package contains also functions to compute Oja signs, Oja signed ranks, Oja ranks, and the related scatter concepts. To illustrate their use, the corresponding multivariate one- and $C$-sample location tests are implemented.

\section*{Keywords}
Oja median, Oja signs, Oja signed ranks, Oja ranks, \proglang{R,C,C++}

\section[Motivation]{Introduction}
The univariate median is a popular location estimator.
It is, however, not straightforward to generalize it to the multivariate
case since no generalization known retains all properties of univariate estimator, and therefore
different generalizations emphasize different properties of the univariate median.
So besides the Oja median described here, there are several other
multivariate median concepts. \citet{Hayford1902} suggested the first generalization
by simply using the vector of the marginal medians.
Other popular multivariate
medians are Tukey's median \citep{Tukey1975} and the spatial median
(also known as $L_1$ median).  The spatial median was initially defined
as a bivariate median \citep{Weber1909,Weber1929} and subsequently extended to
the general multivariate case.
These and more multidimensional medians are surveyed in \cite{Small1990} and \cite{Oja2013}.
While the vector of marginal medians is quite easy
to compute, the other multivariate medians are a more computationally expensive.
Particularly the Oja median \citep{Oja1983} has, despite its compelling statistical properties, not been used very often in practice so far, since it is difficult to compute.
The main topic of this paper is to describe the \pkg{OjaNP} \citep{OjaNP} package, which provides several algorithms for the
computation of the Oja median in \proglang{R} \citep{R}.

The outline of this paper is as follows. In  Section~\ref{subsec:Concepts},
we review and compare some multivariate medians and show which properties
of the univariate median is generalized by which multivariate median.
Our main focus is on the Oja median, whose basic properties are
discussed in Section \ref{subsec:OjaProp}, followed by an introduction to Oja
signs and ranks (Section \ref{subsec:1.2}), Oja signed ranks (Section \ref{subsec:1.4})
and Oja sign and rank covariance matrices (Section \ref{subsec:1.3}).
To demonstrate the application of the Oja median and its sign and rank concepts,
Section \ref{subsec:1.5} discusses one- and $C$-sample tests of location.

In Section \ref{sec:DesAlg}, we focus on the different algorithms provided by the package \pkg{OjaNP} to calculate the Oja median. Four different algorithms are available:
two exact algorithms
and two approximate algorithms based on different designs.

Chapter \ref{sec:Use} shows how to use the package \pkg{OjaNP} in order to calculate
the Oja median and related statistics. We provide simple examples, and additional benchmarks
are calculated to analyze the performance of the implementations.

A concept frequently encountered in this paper is $\emph{affine equivariance}$.
We use it in the sense of full-rank affine equivariance, which is common in robust
statistics. Given the $k$-dimensional sample  $\x_1,\dots,\x_n$  we let
$\X =(\x_1 \dots \x_n)^\top$  be the data matrix of dimension
$n \times k$, containing the data points as rows. Subsequently the
data sample is identified with $\X$. For a given affine-linear transformation
$T: \mathbb{R}^k \to \mathbb{R}^k$, $\bx \mapsto \bo A \bx + \bb$ with $\bb \in \mathbb{R}^k$
and $\bo A \in \mathbb{R}^{k \times k}$ non-singular, the data matrix $\Y$ of the
transformed data $T(\bx_1),\dots,T(\bx_n)$ is given by
\[
	\Y = T(\X) = \X \bo A^\top + \bEins \bb^\top,
\]
where $\bEins$ denotes the $n\times 1$ vector consisting of ones.
We call an $\mathbb{R}^{k}$-valued location statistic
$\bmu(\X)$ \emph{affine equivariant}, if
\begin{equation}\label{affineequi}
\bmu(T(\X)) = T(\bmu(\X)) \quad \text{for all $T$ as above.}
\end{equation}

This applies analogously to set-valued location statistics $\bmu$, such as median sets.
For a matrix-valued scatter statistic $\bS$ taking on values in $\mathbb{R}^{k\times k}$, affine equivariance is commonly understood as
\[
 \bS(T(\X)) = \bo A T(\X) \bo A^\top.
\]

\section{Oja median and related concepts}\label{sec:OMConcepts}
\subsection{Oja median and other multivariate medians} \label{subsec:Concepts}

We start by introducing the univariate median for distributions. Given a distribution
function $F$, let
\begin{eqnarray*}
 F^{-1}\left(\frac{1}{2}-\right)&=&\inf\left\{x \in \mathbb{R}:F(x)\geq \frac{1}{2}\right\} \quad \text{and}\\
 F^{-1}\left(\frac{1}{2}+\right)&=&\sup\left\{x \in \mathbb{R}:F(x)\leq \frac{1}{2}\right\}.
\end{eqnarray*}
Then the \emph{median (set)} of $F$ is given by the interval
\begin{equation}\label{univmedianset}
\text{Med}(F)=\left[F^{-1}\left(\frac{1}{2}-\right)\,,\quad F^{-1}\left(\frac{1}{2}+\right)\right] .
\end{equation}
Any point of the interval divides the distribution in two halves of equal probability weight and can represent the median.
In case a unique selection is needed, we use the gravity center of the median set as a \emph{(single-point) median} and denote it by the lower case symbol,
\begin{equation}\label{univmedianpoint}
 \text{med}(F)=\frac{F^{-1}\left(\frac{1}{2}+\right)+F^{-1}\left(\frac{1}{2}-\right)}{2}.
\end{equation}

For a given sample $\mathbb{X}=(x_1,\dots,x_n)$, the \emph{median}  $\text{med}(\mathbb{X})$ is obtained as
\begin{gather*}
 \text{med}(\mathbb{X})=
\begin{cases}
x_{\left(\frac{n+1}{2}\right)} & \text{if} \; n \; \text{is odd}\,, \\
\frac{1}{2} \left( x_{\left(\frac{n}{2}\right)} + x_{\left(\frac{n}{2}+1\right)}\right) & \text{if} \; n \;\text{is even}\,,
\end{cases}
\end{gather*}
with $x_{(i)}$ being the $i$-th order statistic.
The latter definition is a special case of
(\ref{univmedianpoint}),
obtained by taking $F$ to be the empirical distribution of $\mathbb{X}$, which gives equal probability mass to each of the points
$x_1,\dots,x_n$. Note that $\text{med}(\mathbb{X})$ is an affine equivariant location statistic.

Now let $\X=(\X_1,\ldots,\X_k)$ be a $k$-dimensional dataset, where $\X_i$ denotes the
$i$-th column of $\X$ and corresponds hence to the $i$-th variable.
The many existing notions of a $k$-variate median for such data have in common that they reduce to the univariate median for $k=1$.
Multivariate medians are generally non-unique, and we select, as above, the gravity center of the median set to obtain a unique representation.

To the best of our knowledge the first generalization of the univariate median to
the multivariate case is the vector of marginal medians $\mmed$ described in \citet{Hayford1902}.

\begin{mydef}
 The \emph{vector of marginal
medians} $\mmed$ of the sample $\mathbb{X}$ is defined as
$$
\mmed(\mathbb{X})=(\med(\X_1), \ldots, \med(\X_p))^\top.
$$
\end{mydef}

\begin{figure}[!h]
\centering
\includegraphics[width=0.75\textwidth, height=0.265\textheight]{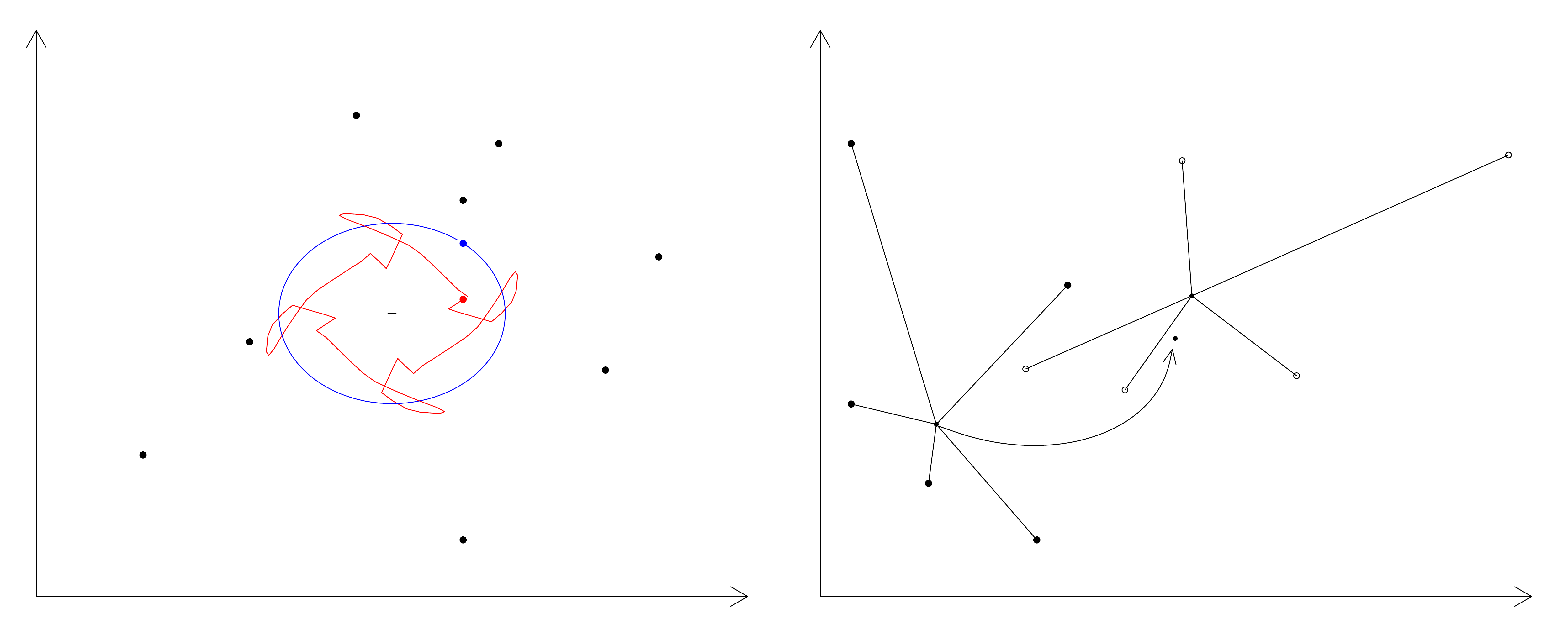}
\caption{Transformation of different multivariate medians.}
\label{fig:compPlot}
\end{figure}

The vector of marginal medians is easily computed but not affine equivariant.
A simple rotation of two-dimensional data visualizes the problem. In the left part of
Figure \ref{fig:compPlot} a two-dimensional dataset is plotted and rotated around the point +.
At the same time the marginal median $\mmed$ is continuously plotted. An affine equivariant
median would follow the rotation on a circle. The blue circle represents the Oja median and
illustrates the behavior of an affine equivariant median. In the right part of
Figure \ref{fig:compPlot} we visualize the behavior of a rotation invariant estimator
that is not scale invariant. We use the spatial median for this example, see Definition \ref{defspatialmedian} below.

The second  generalization of the univariate median reviewed here is based on the fact
that for a given sample $\mathbb{X}$ and its median $\med(x)$ the equation
\[
	\sum_{i=1}^n \id_{(-\infty,\med(x)]}(x_i)=\sum_{i=1}^n \id_{[\med(x),\infty)}(x_i)
\]
holds, where $\id_{A}(x_i)=1$ if $x_i \in A$ and $= 0$ otherwise. This means,
there are as many observations smaller than $\med(x)$ as are bigger, as it
was pointed out e.g.\ in \citet{Hotelling1929}.
Given a $k$-dimensional dataset, we consider halfspaces in every direction $\mathbf{v}\in S^{k-1} = \{ \bx\ |\ ||\bx|| = 1 \}$.
Let $H_\mathbf{v}$ denote the ``minimal'' halfspace with normal vector $\mathbf{v}$ that contains at least half of the data points, i.e., for any other halfspace $\tilde{H}_\mathbf{v}$ with these properties $H_\mathbf{v} \subset \tilde{H}_\mathbf{v}$ holds.
The intersection of all $H_\mathbf{v}, \mathbf{v}\in S^{k-1}$, forms the \emph{Tukey median} {\bf Tmed}($\mathbb{X}$).
If the data are in general position, each such halfspace is bordered by a hyperplane through exactly $k$ data points, and the Tukey median is in general no singleton.
(A set of $k$-variate data is in \emph{general position} if at most $k$ of them lie on the same hyperplane.)
The unique \emph{Tukey median} {\bf tmed}($\mathbb{X}$) is defined as the gravity point of this median set; see \citet{Tukey1975} and \citet{Donoho1992}.
It is affine equivariant \citep{Chen1995} and can be introduced as the maximizer of a depth function as follows.

\begin{mydef}
Let $\X$ be a $k$-dimensional sample as above. For any $\mathbf{x}\in \mathbb{R}^k$,
\begin{equation}\label{Tukeydepth}
\text{\bf tdep}_\X(\bx)= \frac 1n \min_{\|\mathbf{v}\|=1}\#\{i : \mathbf{v}^\top\mathbf{x}_i\ge \mathbf{v}^\top\mathbf{x} \}
\end{equation}
is called the \emph{Tukey depth} or \emph{location depth} of $\mathbf{x}$ w.r.t.\ $\mathbb{X}$. (Here $\#\{S\}$ denotes the cardinality of a set $S$.)
The \emph{Tukey median} {\bf Tmed}$(\mathbb{X})$ of the sample $\mathbb{X}$ is defined as the maximizer of the Tukey depth,
\begin{equation}
\text{\bf Tmed}(\mathbb{X}) = \argmax_{\mathbf{x}\in\mathbb{R}^k}
\{\mathbf{tdep}_\X(\mathbf{x})\}\,.
\end{equation}
\end{mydef}

Another way to generalize the univariate median is to transfer its minimizing feature into
higher dimensions. Consider again a univariate sample $\mathbb{X}=(x_1,\dots,x_n)$ in $\mathbb{R}$. When minimizing the sum
$ \sum_{i=1}^n |x_i - x|$ over $x\in \mathbb{R}$ we obtain
\begin{equation}\label{univariate.objective}
\text{med}(\mathbb{X})= \argmin_{x\in\R} \sum_{i=1}^n |x_i - x|\,.
\end{equation}
This minimizing feature of the univariate median is interpretable in two ways: it is the sum of absolute deviations,
but it can also be viewed as the sum of one-dimensional simplices. The first
interpretation will lead us to the spatial median, the second to the Oja median.

The spatial median is presumably the most popular multivariate median and almost as old
as the marginal median. \citet{Weber1909} first described the spatial median and used
it to solve an economic problem: he was looking for the best place of a
distribution center in the sense, that the sum of distances between outposts
and the distribution center becomes minimal. The $k$-dimensional extension of this approach
is the spatial median.

\begin{mydef}\label{defspatialmedian}
The spatial median $\smed$ of a $k$-variate data sample $\X$ is defined as
\be \label{spatial.objective}
	\smed(\X) =
	\argmin_{ \mathbf{x}\in \mathbb{R}^k}
	\left\{ \sum_{i=1}^n \|\mathbf{x}_i-\mathbf{x}\|_2\right\}.
\ee	
\end{mydef}
Here $\|\cdot \|_2$ denotes the Euclidean norm. The spatial median is sometimes also referred to as the \emph{$L_1$ median} since it minimizes the $L_1$ norm of the $n$-variate vector of distances $\|\mathbf{x}_i-\mathbf{x}\|_2$.
The spatial median is not affine equivariant as is visualized in the right part of Figure \ref{fig:compPlot}.
Given an affine transformation
$T:\mathbb{R}^2 \rightarrow \mathbb{R}^2$ which transforms the data points from
the left star-shaped figure into the right star-shaped figure, the center of each star is
the spatial median of the data points.
However, due to the lack of affine equivariance, the transformed spatial median $T(\smed(\X))$ (marked with the arrow) does not coincide with the spatial median $\smed(T(\X))$ of the transformed data set. The spatial median is rotation invariant, but not scale invariant.

\citet{Oja1983} introduced a multivariate median based on volumes of simplices. A $k$-dimensional simplex is the
convex hull of $(k+1)$ spanning points in general location. Let
$$
\mathbf{x}_1=(x_{1,1},\dots,x_{1,k})^\top,
\mathbf{x}_2=(x_{2,1},\dots,x_{2,k})^\top,
\dots,
\mathbf{x}_{k+1}=(x_{k+1,1},\dots,x_{k+1,k})^\top
$$
be $k+1$ points in general location from $\mathbb{R}^k$. The volume $V(\mathbf{x}_1,\dots,\mathbf{x}_{k+1})$ of the simplex
spanned by the points $\mathbf{x}_1,\dots,\mathbf{x}_{k+1}$ is then given by
\begin{eqnarray}\label{volumenOja}
 V(\mathbf{x}_1,\dots,\mathbf{x}_{k+1})=\abs\left(\frac{1}{k!}\det
\begin{pmatrix}
1       & 1       & \cdots & 1\\
x_{1,1} & x_{2,1} & \cdots & x_{k+1,1}\\
x_{1,2} & x_{2,2} & \cdots & x_{k+1,2}\\
\vdots  & \vdots  &        & \vdots \\
x_{1,k} & x_{2,k} & \cdots & x_{k+1,k}
\end{pmatrix}
\right) \ ,
\end{eqnarray}
see, e.g., \citet{Stein1966}. Let $F$ be a distribution on $\mathbb{R}^k$ having finite first moment.
The Oja median $\text{\bf Omed}(F)$ of $F$ is defined as follows:
for i.i.d.\ random vectors $\mathbf{X}_1,\dots,\mathbf{X}_k$, each distributed  with $F$, define the Oja median set as
\begin{eqnarray*}
\text{\bf Omed}(F)= \underset{\boldsymbol{\mu} \in \mathbb{R}^k}{\argmin} \; E(V(\mathbf{X}_1,\dots,\mathbf{X}_k,\boldsymbol{\mu})).
\end{eqnarray*}
For data we have the following definition. Let
\be \label{Pnk}
	P_{n,k} \ = \ \left\{\, p=(i_1,\dots,i_k) \,\middle|\, 1 \leq i_1 < \dots < i_k \leq n \right \}
\ee
be the set of all ordered $k$-tuples out of $\{1,\dots,n\}$, $1 \le k \le n$.

\begin{mydef}
The Oja median $\text{\bf Omed}$ of a $k$-dimensional sample $\mathbb{X}=(\mathbf{x}_1 \dots \mathbf{x}_n)^\top$
of size $n > k$ is defined as
\be \label{Oja.objective}
	\text{\bf Omed}(\X)=
	\underset{\mathbf{x} \in \mathbb{R}^k}{\mbox{\rm argmin}}
	\left\{ \sum_{(i_1,\dots,i_k) \in P_{n,k}}V(\mathbf{x}_{i_1},\dots,\mathbf{x}_{i_k},\mathbf{x}) \right\}.
\ee
\end{mydef}
This means that the Oja median of a $k$-dimensional sample is any point $\mathbf{x} \in \mathbb{R}^k$ for which
the sum of simplex volumes over all combinations of possible $k$ data points is minimal. Note that $\text{\bf Omed}(\X)$ equals $\text{\bf Omed}(F_\X)$, where $F_\X$ is the empirical distribution on $\mathbf{x}_1 \dots \mathbf{x}_n$. A unique version of the Oja median, denoted $\text{\bf omed}(\X)$, is obtained by selecting the point of gravity of $\text{\bf Omed}(\X)$.

\subsection{Properties of the Oja median} \label{subsec:OjaProp}

As other multivariate extensions of the median, the Oja median is not unique. In the bivariate case, we have the following result.

\begin{mytheorem}
If $n$ is even and the data points $\mathbf{x}_1,\ldots,\mathbf{x}_n \in \mathbb{R}^2$ are in general position, then the Oja median is unique.
\end{mytheorem}

\begin{figure}[ht]
\centering
\includegraphics[width=\textwidth]{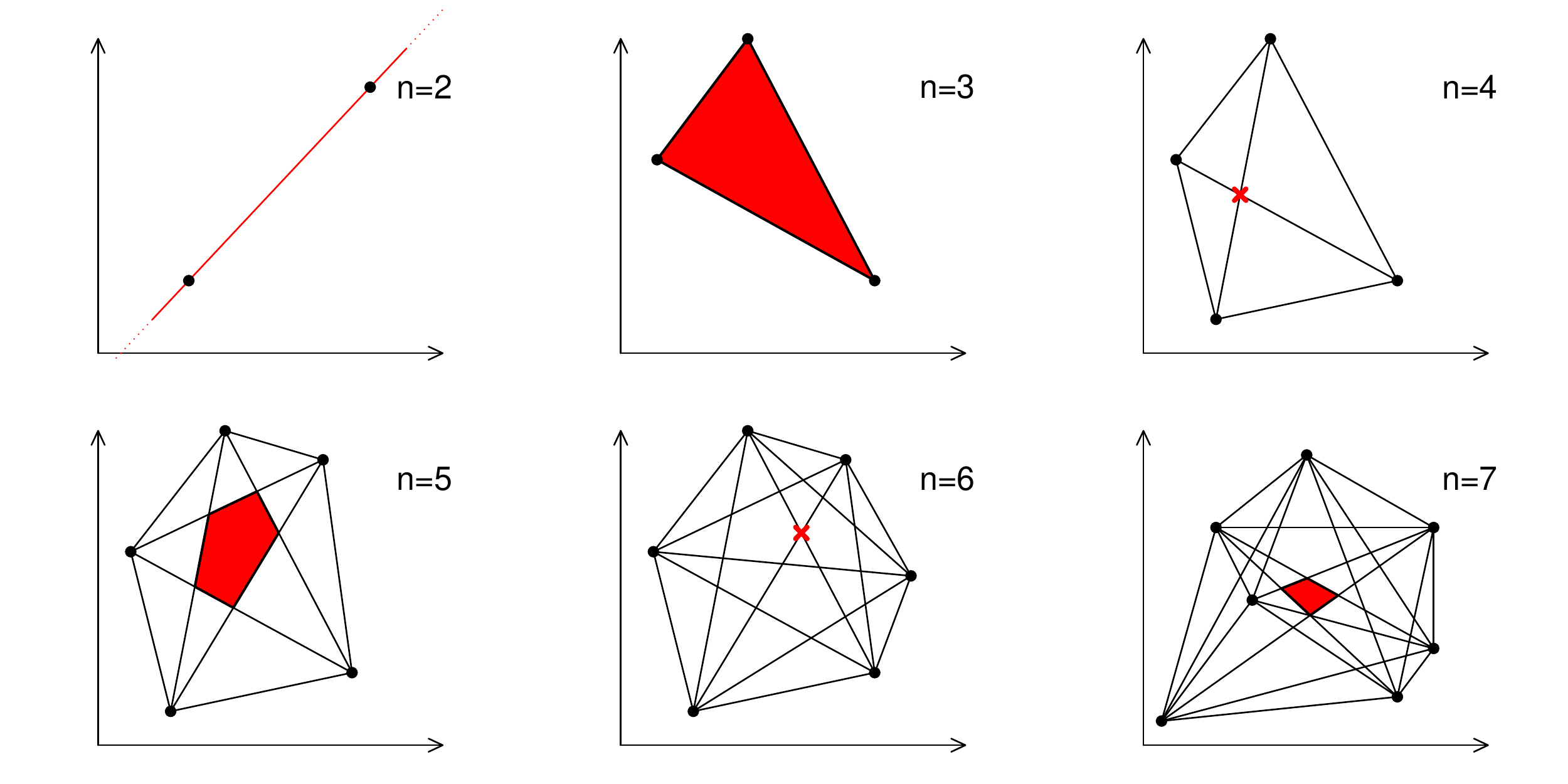}
\caption{Example plots for the bivariate case}
\label{fig:examplePlot}
\end{figure}

For details see \citet{Niinimaa1995}. The author also identifies a necessary and sufficient condition for the bivariate Oja median to be unique if $n$ is odd. There appears to be no result for higher dimensions, but \citet{Oja1999} conjectures that the Oja median is unique for even (odd) sample sizes if the dimension $k$ is even (odd). Figure \ref{fig:examplePlot} visualizes the Oja Median in several small-sample data situations.

\begin{figure}[ht]
\centering
\includegraphics[width=\textwidth]{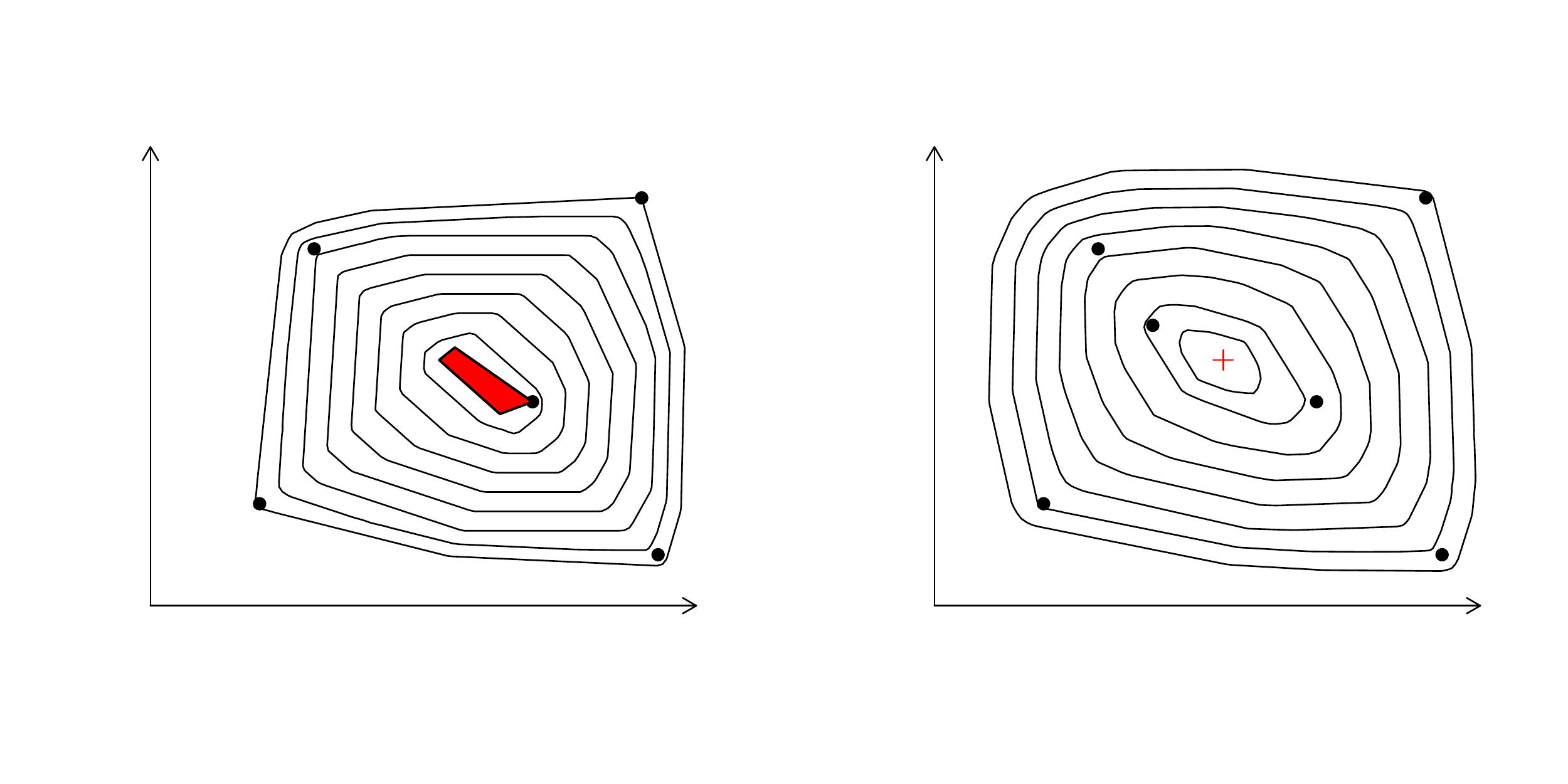}
\caption{Contour plot for the bivariate case.}
\label{fig:contourPlot}
\end{figure}

\begin{mytheorem}
\label{th:omConvex}
The Oja median of a sample is a convex set.
\end{mytheorem}
See \cite{Oja1985} for details. Theorem \ref{th:omConvex} is illustrated in Figure~\ref{fig:contourPlot}.
Contour lines of the objective function (\ref{Oja.objective}) are plotted for two small data clouds. The objective function is convex for, both, even and odd sample sizes.

\begin{mytheorem}
\label{th:ae}
The Oja median is affine equivariant.
\end{mytheorem}

Hence the Oja median is proper location statistic in the sense of (\ref{affineequi}), i.e., $\text{\bf omed}(T(\X)) = T(\text{\bf omed}(\X))$. The next result can be found, e.g., in \citet{ArconesChenGine1994}, \citet{Shen2008}.

\begin{mytheorem}
Under mild regularity conditions (including the existence of first moments) for an i.i.d.\ sample $\X = (\bo x_1 \ldots \bo x_n)^\top$ from the distribution $F$, we have
\[
\sqrt n (\text{\bf omed}(\X)-\text{\bf omed}(F)) \rightarrow_d  N_k(\bo 0, \bo A^{-1} \bo B \bo ({\bo A}^{-1})^\top),
\]
where $\bo B$ is the Oja sign covariance matrix ($\OSCM$) defined later in Section~\ref{subsec:1.3} at $F$ and $\bo A$ is the expected covariance matrix between the Oja signs ($\osgn$, see Section~\ref{subsec:1.2})
and the optimal location score of $F$.
\end{mytheorem}

Concerning robustness the Oja Median has the following properties.
\begin{mytheorem}
Let $F$ have a finite first moment. Then the Oja median has a bounded influence function.
\end{mytheorem}

The influence function is for example given in \cite{Niinimaa1995b}. The Oja median has an asymptotic breakdown point of 0.
\citet{Niinimaa1990} show the following:
\begin{mytheorem}
The finite-sample breakdown-point of the bivariate Oja median is $2/(n+2)$.
\end{mytheorem}
Table~\ref{table:mainProp} summarizes the main properties of the Oja median and
other common multivariate medians discussed here. For another recent discussion of the different medians see also \citet{Oja2013}.

\renewcommand{\arraystretch}{1.1}
\begin{table}[ht]
\caption{Main properties of different multivariate medians}
\centering
\begin{tabular}{c c c}
\hline\hline
Median       & Affine Equivariance & Breakdown Point\\ [0.5ex]
\hline
Marginal Median    & No   & $1/2$                    \\
Tukey Median & Yes  & $1/(k+1)$ \\
Spatial Median      & No  & $1/2$       \\
Oja Median   & Yes  & 0               \\

\hline\hline
             & Influence function & Asymptotic distribution \\ [0.5ex]
\hline
Marginal Medians    & \cite{Niinimaa1995b} & \cite{Babu1988}\\
Tukey Median & \cite{Romanazzi2001} & \cite{Bai1999}\\
Spatial Median     & \cite{Niinimaa1995b} & \cite{Mottonen2010}\\
Oja Median   & \cite{Niinimaa1995b} & \cite{Shen2008} \\
\end{tabular}
\label{table:mainProp}
\end{table}

\subsection{Oja signs and ranks} \label{subsec:1.2}

Closely related to the median is the concept of signs and ranks. In this subsection we introduce
the multivariate \emph{Oja sign} and \emph{Oja rank} and relate them to other multivariate signs
and ranks, corresponding to the marginal and the spatial median. Again, to motivate the subsequent
derivations we take a brief look at the univariate case. Suppose we have a univariate data set
$\mathbb{X} = (x_1 \dots x_n)^\top$, $n \in \mathbb{N}$. We call
\be \label{sign}
	\sgnx(x) = \sgn(x - \med(\X)), \quad x \in \mathbb{R},
\ee
the \emph{sign of $x$ w.r.t.\ the data sample $\mathbb{X}$}, where $\sgn$ is the univariate sign
function ($\mathrm{sgn}(x) = \frac{x}{|\,x|}$ if $x \ne 0$ and zero otherwise), and
$\mathrm{med}(\mathbb{X})$ is the univariate median of the sample $\mathbb{X}$. There are several
possibilities of suitably assigning ranks to the data points. By
\be \label{rank}
	\rnkx(x) = \avei \sgn(x - x_i), \quad x \in \mathbb{R},
\ee
we define normalized central ranks, which may take on $2 n +1$ possible values ranging from -1 to 1.
In the following we call $\rnkx(x)$ simply the \emph{rank of $x$ w.r.t.\ $\X$}.
\par
The median appropriately centers the data. I.e.,\ the signs of the data points, centered by the median,
sum up to zero:
\be \label{sign-med}
	\sumi \sgnx(x_i) = \sumi \sgn(x_i- \med(\X)) = 0.
\ee
(Note that the mean does the same for the data points themselves.) In other words we may say that the median
has central rank zero,
\be \label{rank-med}
	\rnkx(\med(\X)) = - \avei \sgn(x_i- \med(\X)) = 0\,,
\ee
and, in this respect, is the \emph{most central point}.
Identities (\ref{sign-med}) and (\ref{rank-med}) (which are different formulations of the fact that
half of the data lies above and below the median) provide the essential link between signs and ranks
and the median and are a motivating principle behind the multivariate sign and rank functions we will
introduce next. Note that a $k$-variate sign function should be a vector that can point in any direction of the $k$-dimensional space.
The same holds for a multivariate rank function based on signs.

An obvious extension of (\ref{sign}) to the multivariate setting is its componentwise
application, leading to the \emph{marginal sign} function. We call
\[
 	\msgnx(\x) = \msgn(\x - \mmed(\X))
\]
the \emph{marginal sign of $\bx \in \mathbb{R}^k$ w.r.t.\ the $k$-variate data sample
$\X = (\bx_1 \ldots \bx_n)^\top$}, where $\x = (x_1 \ldots x_k)^\top$,
\[
	\msgn(\x) = ( \sgn(x_1) \dots \sgn(x_k) )^\top,
\]
and $\mmed(\X)$ is the marginal median of $\X$. An equally straightforward generalization is the
\emph{spatial sign of $\x$ w.r.t.\ $\X$}:
\[
	\ssgnx(\x) = \ssgn(\x - \smed(\X)), \quad  \x \in \mathbb{R}^k,
\]
where
\[
	\ssgn(\x) =
		\begin{cases}
  				\frac{1}{||\,\x||} \ \x & \mbox{ if } \x \ne \bNull, \\
  				\bNull					 	  & \mbox{ if } \x = \bNull
  	\end{cases}
\]
and $\smed(\X)$ is the spatial median of $\X$.
The corresponding rank functions, the \emph{marginal rank} $\mrnkx$ and the \emph{spatial rank} $\srnkx$,
are obtained by replacing $\sgn$ in (\ref{rank}) by $\msgn$ and $\ssgn$, respectively.
The \emph{Oja sign} is defined as follows. For $0 \le k \le n$,
let $N_{n,k} =
\binom{n}{k}$ and $P_{n,k}$ as in $(\ref{Pnk})$.
We call
\begin{eqnarray} \label{osgnxm}
	\mbox{\bf osgn}_{\mathbb{X}}(\bx; \bm)
	& = &
	\frac{1}{N_{n,k-1}}
	\sum\limits_{(i_1,\ldots,i_{k-1})\in P_{n,k-1}} \nabla_{\x}
	\left| \,
		\det(\x_{i_1}-\bm,\dots,\x_{i_{k-1}}-\bm ,\x-\bm)
	\right| \\
	& = &
	\frac{1}{N_{n,k-1}}\sum\limits_{(i_1,\ldots,i_{k-1})\in P_{n,k-1}} \nabla_{\x}
	\left| \, \det
	 \left( \begin{matrix}
			1 	&  1  & \dots & 1 & 1 \\
			\bm 	& \x_{i_1} & \dots & \x_{i_{k-1}} & \x
			\end{matrix}
	\right)
	\right| \nonumber
\end{eqnarray}
the \emph{Oja sign of the point $\x \in \mathbb{R}^k$ w.r.t.\ the data sample $\mathbb{X}$ and the center location $\bm$} and
\be \label{osgnx}
	\osgnx(\x) = \osgnx(\x ; \omed(\X))
\ee
simply the \emph{Oja sign of $\x$ w.r.t.\ $\X$}. Furthermore
\begin{eqnarray}
	\ornkx(\x) & = & \frac{1}{n-k+1} \sum_{i=1}^n \osgnx(\x ; \x_i)  \nonumber \\
	           & = & \frac{1}{N_{n,k}}
	           \sum\limits_{(i_1,\ldots,i_{k})\in P_{n,k}} \nabla_{\x}
	                	\left| \, \det
	           			\left( \begin{matrix}
										1 			 & \dots & 1 			  & 1 \\
										\x_{i_1} & \dots & \x_{i_k} & \x
									\end{matrix} \right)
										\right| \label{ornkx}
\end{eqnarray}
is the \emph{Oja rank of $\x$ w.r.t.\ $\X$}.
The notation $\nabla_{\x}$ means the gradient (the derivative as a column vector) w.r.t.\ $\x$. We
define the derivative of $|\cdot|$ to be zero at the origin, i.e.,\
$\frac{d}{d x}|x| = \sgn(x)$, $x \in \mathbb{R}$.

We note a qualitative difference in the definition of the Oja sign to the marginal and the spatial sign.
The Oja sign function $\osgnx$ depends on the sample $\X$ \emph{not only} through the centering point
$\omed(\X)$, but is a function of the whole sample. By introducing the parameter $\bm$ in definition
(\ref{osgnxm}) we allow the data to be centered by a different central location than the Oja median.
This  may be useful under some circumstances, for example to speed up computation. The function
\code{ojaSign} offers this option.

We also note that for $k =1$ expressions (\ref{osgnx}) and (\ref{ornkx}) reduce to
\[
	\osgnx(x) = \frac{d}{d x} \left|\, \det( x - \med(\X)) \, \right| = \sgnx(x)
\]
and
\[
	\ornkx(x) = \avei \frac{d}{d x}
	   	\left| \, \det
	           			\left( \begin{matrix}
										1   & 1  \\
										x_i & x
									\end{matrix} \right)
										\right|
	= \avei \sgn(x - x_i) = \rnkx(x),
\]
hence $\osgnx$ and $\ornkx$ are proper generalizations of $\sgnx$ and $\rnkx$, respectively, in the sense
that for $k = 1$ they coincide with their univariate counterparts.

In dimensions 2 and 3 Oja signs and ranks allow a descriptive geometric interpretation. The key is to
recall that the $k$-dimensional volume of the simplex spanned by $k+1$ points $\x_1,\dots,\x_{k+1}$ in
$\mathbb{R}^k$ is given by $V(\bx_1,\dots,\bx_{k+1})$, cf.\ (\ref{volumenOja}), and that hence
\[
	\det
	           			\left( \begin{matrix}
										1 			 & \dots & 1 			  & 1 \\
										\x_{1} & \dots & \x_{k} & \x
									\end{matrix} \right)
									= 0
\]
characterizes all points $\x \in \mathbb{R}^k$ that lie on the $k-1$ dimensional hyperplane spanned by the $k$
points $\x_1,\dots,\x_k$. Then, for $k = 3$ ($k = 2$), $\osgnx(\x)$ is the average of the
$N_{n,k-1} = \binom{n}{k-1}$ $k$-dimensional vectors $\bv_p$, $p = (i_1,\dots, i_{k-1}) \in P_{n,k-1}$,
where $\bv_p$
\bit
\item
is perpendicular to the plane (straight line) going through the $k$ points $\x_{i_1},\dots,\x_{i_{k-1}}$ and $\omed(\X)$,
\item
has length equal to $(k-1)!$ times the area (length) of the triangle (line segment) that is bordered by the $k$ points and
\item
points from the plane (straight line) to the point $\x$.
\eit
If $\x$ itself lies on the plane (straight line) or if the $k$ points do not uniquely determine a plane
(straight line), the vector $\bv_p$ is zero. Likewise, $\ornkx(\x)$ is the average of
$N_{n,k} = \binom{n}{k}$ vectors $\bv_p, p = (i_1, \dots, i_k) \in P_{n,k}$, each perpendicular to the
the plane (straight line) spanned by $\x_{i_1}, \dots, \x_{i_k}$.
\begin{figure}[t]
\centering
\includegraphics[width=0.4\linewidth]{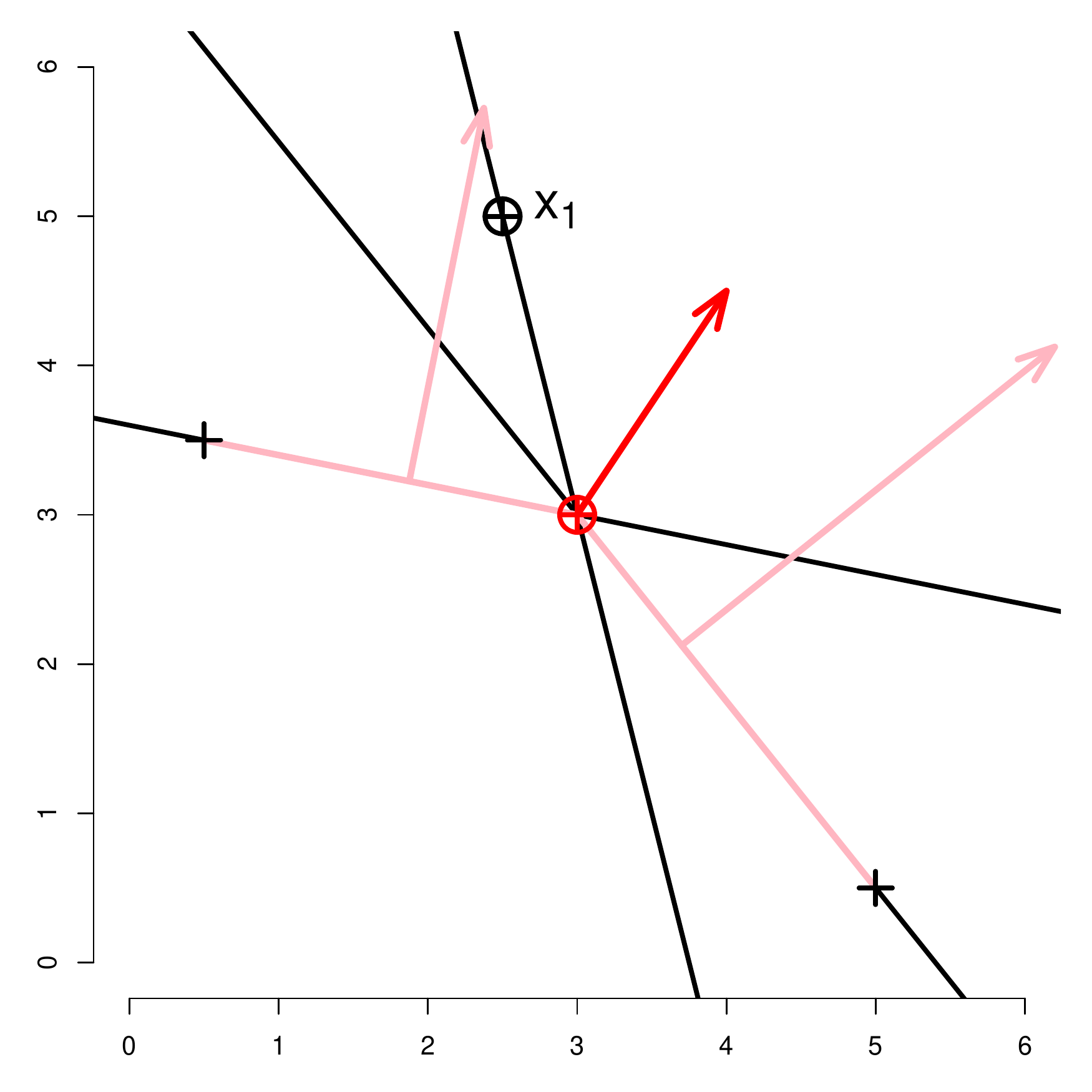}
\caption{
		Three points in $\mathbb{R}^2$: The Oja sign (in red) of point $\x_1$ is the average of three vectors: the light
red vectors and the null vector. }
\label{Fig:Oja_sign}
\end{figure}
The construction of the Oja sign is visualized in Figure \ref{Fig:Oja_sign} at the very simple example of three
points in $\mathbb{R}^2$: The red $\oplus$ is an interior point (let it be denoted by $\bm_0$) of the Oja median set, which
is simply the triangle bordered by the three data points. Then $\osgnx(\bx_1;\bm_0)$, the vector in red, is formed
as the average of the two light red vectors and the null vector (the latter being interpreted as a vector pointing perpendicularly
from the straight line $\bx_1$---$\bm_0$ to $\bx_1$). In this example, $\osgnx(\bx_1;\bm_0)$ is invariant w.r.t.\
the specific choice of $\bm_0$ as long as $\bm_0$ is below both lines $\x_1$---$\x_2$ and $\x_1$---$\x_3$.

As in the univariate case, cf.\ (\ref{sign-med}) and (\ref{rank-med}), the multivariate signs of the data points sum up to zero, i.e.,\
\[
		\sumi \bsgnx(x_i) =  \bNull,
\]
where $\bsgnx$ may be any of $\msgnx$, $\ssgnx$ or $\osgnx$. Equivalently,
\be \label{multivariate.rank-med}
		\brnkx(\bmed(\X)) = \bNull,
\ee
where $\brnkx$ may be any of $\mrnkx$, $\srnkx$ or $\ornkx$, and $\bmed(\X)$ the respective multivariate median. This
is generally only true if the appropriate median is chosen as centering point, for Oja signs in particular, the data
has to be centered by the Oja median. Some further care must be given to these statements. If the median set is no singleton,
they may be false for median points on its border. This also happens in the univariate case, when, for an even
number $n$ of observations, one chooses the median to be $x_{(\frac{n}{2})}$ or $x_{(\frac{n}{2}+1)}$. The function
\code{ojaMedian} is likely to return points at the border of the median set. In (\ref{spatial.objective}) and (\ref{Oja.objective}), respectively, the spatial median and the Oja median were introduced as minimizers of objective functions,
both generalizing the univariate case (\ref{univariate.objective}). Spatial and Oja ranks may be introduced as derivatives of these
objective functions. Thus (\ref{multivariate.rank-med}) is in concordance with the optimality property of the medians.

The negative rank function $-\ornkx(\x)$ defines a hyperplane through $\bo{x}$, on the positive side of which the Oja median is found. This property is used by the exact bounded algorithm (Section \ref{sec:eba}) to reduce the search region for the median.

Similar to the multivariate medians, the multivariate signs differ by their equivariance and invariance properties. All multivariate
signs are \emph{translation invariant}, i.e.,\ for
$\Y = \X + \bEins_n \bb^\top$, $\bb \in \mathbb{R}^k$, we have
\[
	 {\bsgn}_{\Y}(\bx + \bb) = \bsgnx(\bx), \qquad \bx \in \mathbb{R}^k,
\]
where, again, $\bsgn$ is any of $\msgn$, $\ssgn$, $\osgn$. Marginal signs are furthermore invariant w.r.t.\ monotonously
increasing, componentwise transformations, in particular
\[
	{\msgn}_{\X \bo D ^\top}(\bo D \bx) = \msgnx(\bx), \qquad \bx \in \mathbb{R}^k,
\]
for $\bo D = \diag(d_1,\dots,d_k)$ with $d_i > 0$, $i = 1,\dots,k$. Spatial signs on the other hand are equivariant under
orthogonal transformations, i.e.,\
\[
	\ssgn\nolimits_{\X \bo U^\top}(\bo U \bx) = \bo U \, \ssgnx(\bx), \qquad \bx \in \mathbb{R}^k,
\]
for any orthogonal $\bo U \in \mathbb{R}^{k \times k}$. Oja signs even obey a form of affine equivariance, an \emph{inverse proportional affine equivariance}:
\be \label{sign.affine}
	\osgn\nolimits_{\X \bo A^\top}(\bo A\x) = \det(\bo A) (\bo A^{-1})^\top \osgnx(\x), \qquad \bx \in \mathbb{R}^k,
\ee
for any non-singular $\bo A \in \mathbb{R}^{k \times k}$, and they appear in the literature under the name \emph{affine equivariant signs}.
The respective multivariate rank functions have analogous equivariance and invariance properties. More details on Oja signs and
ranks and their applications can be found, e.g., in \citet{Oja1999}. \citet{HettmanspergerMcKean2011} give an overview of
multivariate sign and rank methods in general, \citet{PuriSen1971} treat methods based on marginal signs and ranks and \citet{Oja2010}
describes spatial sign and rank methods.

\subsection{Oja signed ranks} \label{subsec:1.4}

As in the Section \ref{subsec:1.2} we consider first the univariate case.  Suppose
we have a univariate data set $\X=(x_1,\ldots, x_n)$.
The \emph{signed rank of $x$ w.r.t.\ $\X$} can be defined by
\begin{eqnarray}
\label{signedrank}
\rnkxb(x) & = & \frac{1}{2n}\sum_{i=1}^n\sum_{a\in A}\sgn(x-a x_i),
\end{eqnarray}
where $A=\lbrace -1,1\rbrace$. Note that the signed rank
of the $j$-th observation can be written as
\begin{eqnarray}
\rnkxb(x_j) & = & \frac{1}{2n}\lbrace 2\,\rnk(|x_j|)-1\rbrace\sgn(x_j),
\end{eqnarray}
where $|x_j|$ is ranked among absolute values $|x_1|,\ldots,|x_n|$.
The Wilcoxon signed rank statistic
$\sum_{i=1}^n\rnk(|x_i|)\sgn(x_i)$ is thus asymptotically
equivalent with $n\sum_{i=1}^n\rnkxb(x_i)$.

The signed rank can be extended to the multivariate case, provided we have a concept of
multivariate sign. We get \emph{spatial signed rank} $\srnkxb(\x)$ if we replace $\sgn$ in (\ref{signedrank}) by
$\ssgn$. The \emph{Oja signed rank} can be defined in the following way.
Let $\mathbb{A}$ be the set of $2^k$ possible vectors $(\pm 1,\ldots,\pm 1)$, i.e.,
$\mathbb{A} = \lbrace \mathbf{a}=(a_1,\ldots,a_k) : a_1=\pm 1, \ldots, a_k=\pm 1\rbrace$. Then
\begin{eqnarray}
	\ornkxb(\x)
	           & = & \frac{1}{2^kN_{n,k}}\sum\limits_{P_{n,k}}
	                    \sum_{\mathbf{a}\in \mathbb{A}}
	                    \nabla_{\x}
	                	\left| \, \det
	           			\left( \begin{matrix}
						   1 & \cdots & 1 & 1  \\
					       a_1\x_{i_1} & \cdots & a_k\x_{i_k} & \x
						\end{matrix} \right)
						\right| \label{ornkxb}
\end{eqnarray}
is the  \emph{Oja signed rank of $\x$ w.r.t.\ $\X$}. It is easy to see that the Oja signed rank coincides
with the univariate signed rank when $k=1$. Furthermore, it can be shown that
\begin{itemize}
\item
$\ornkxb(\x)$ is odd: $\ornkxb(-\x) = -\ornkxb(\x)$.
\item
$\ornkxb(\x)$ points (approximately) in the direction of $\x$.
\item
$\ornkxb(\bNull)=\bNull$.
\item
$\ornkxb(\x)$ is bounded, piecewise constant and increases in magnitude as $\x$ moves away from
$\bNull$.
\end{itemize}

\subsection{Oja sign and rank matrices} \label{subsec:1.3}

\begin{figure}[t]
\begin{minipage}{0.50\linewidth}
{\bf a)}
\includegraphics[width=0.8\linewidth]{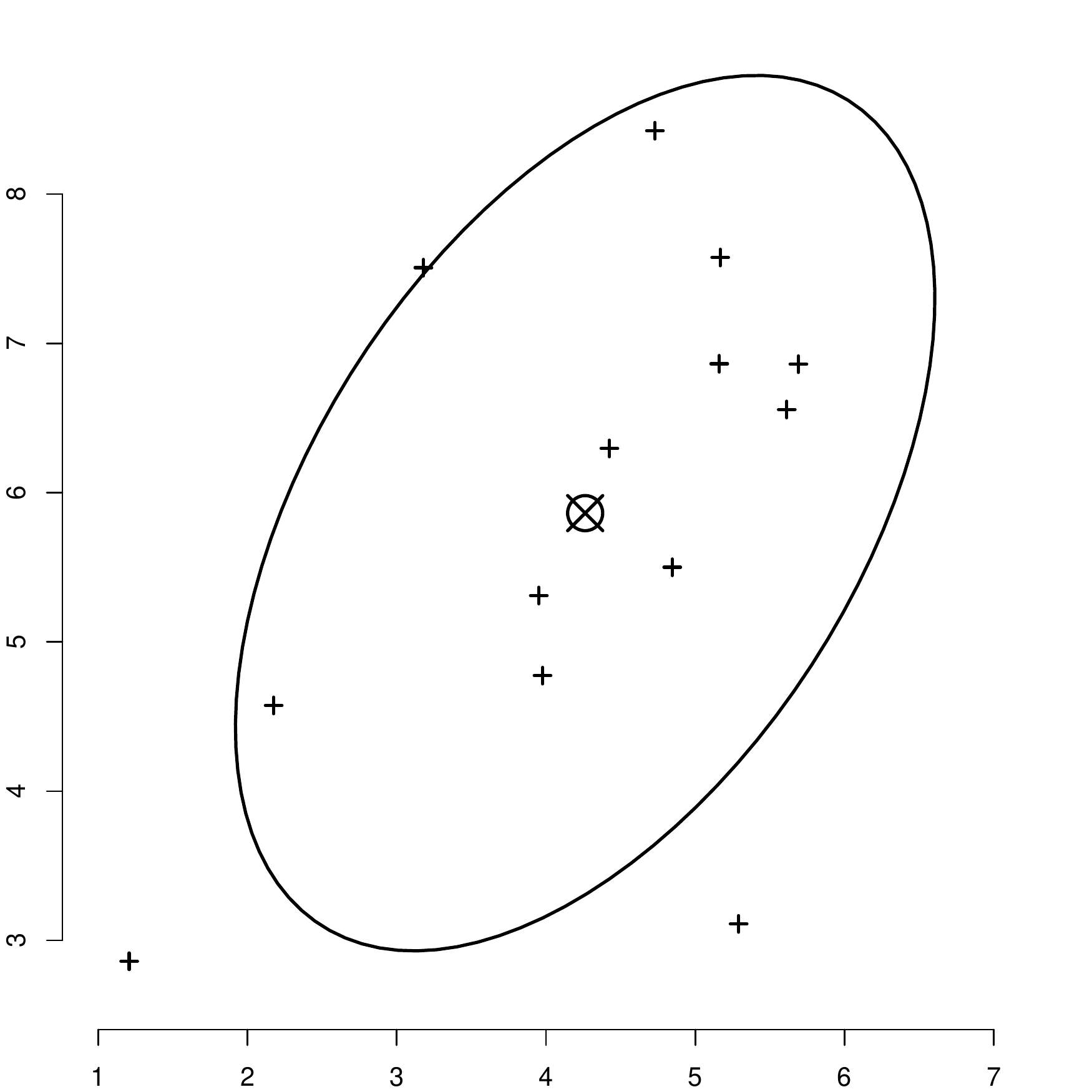}\\
{\bf b)}
\includegraphics[width=0.8\linewidth]{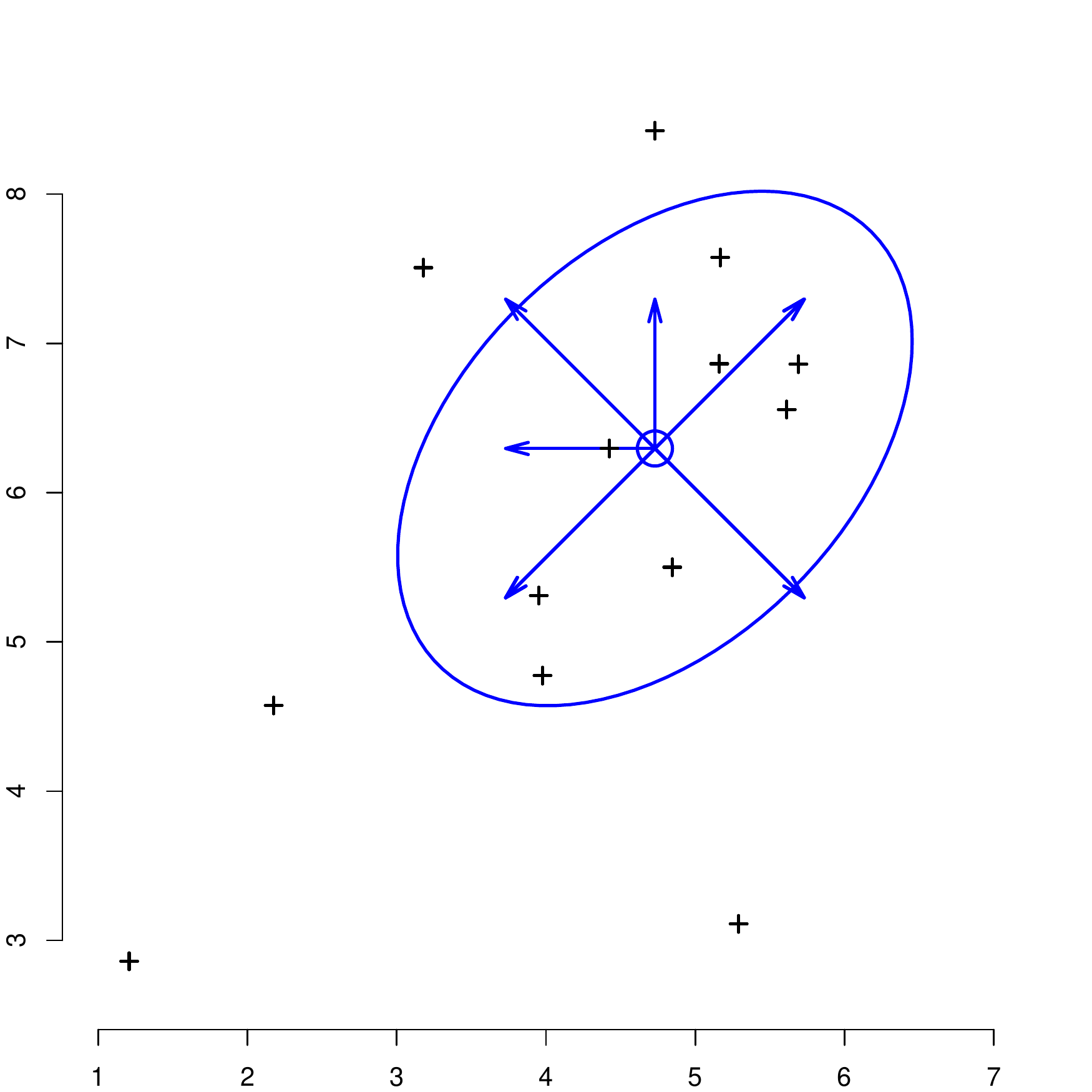}
\smallskip
\end{minipage}
\hfill
\begin{minipage}{0.50\linewidth}
{\bf c)} 
\includegraphics[width=0.8\linewidth]{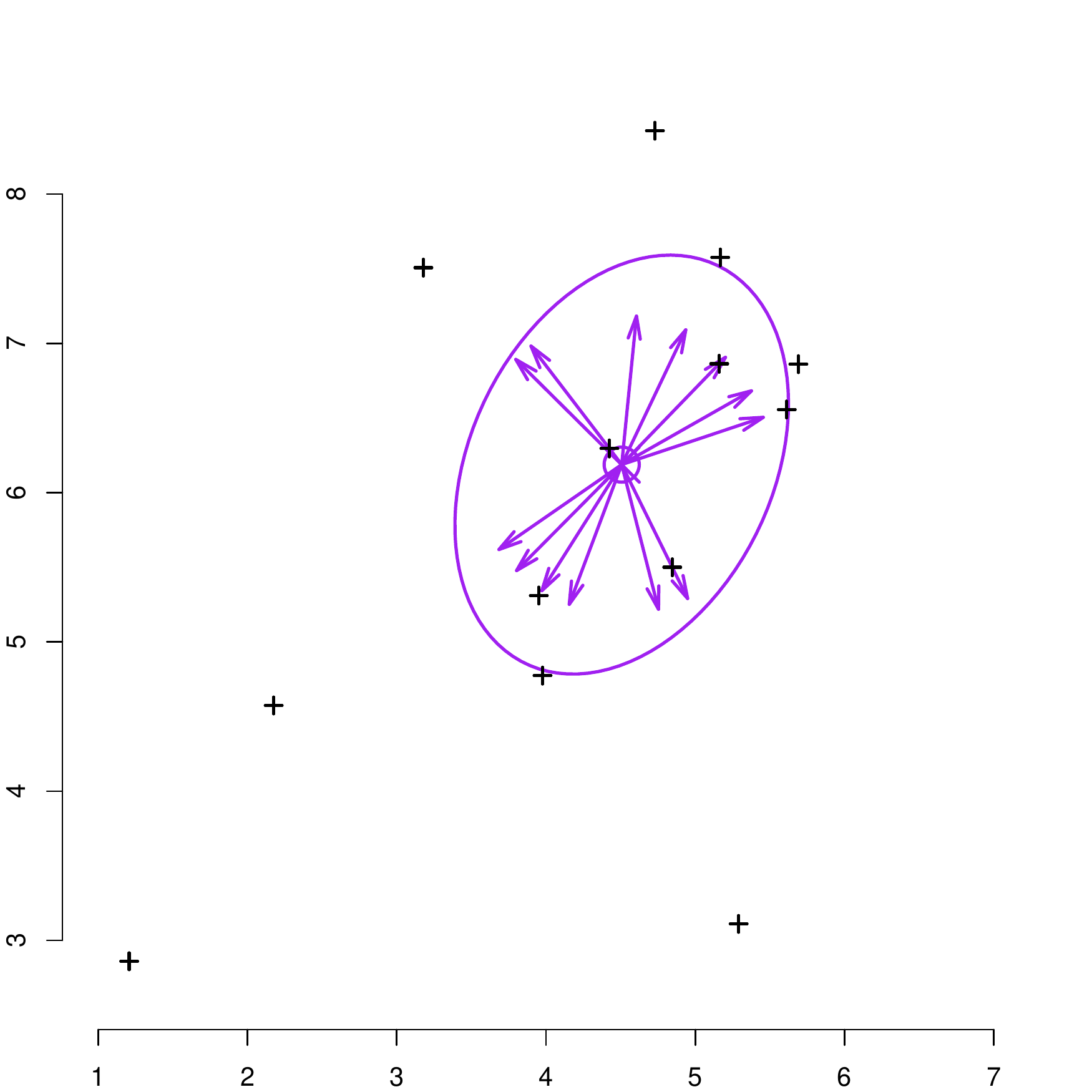}\\
{\bf d)} 
\includegraphics[width=0.8\linewidth]{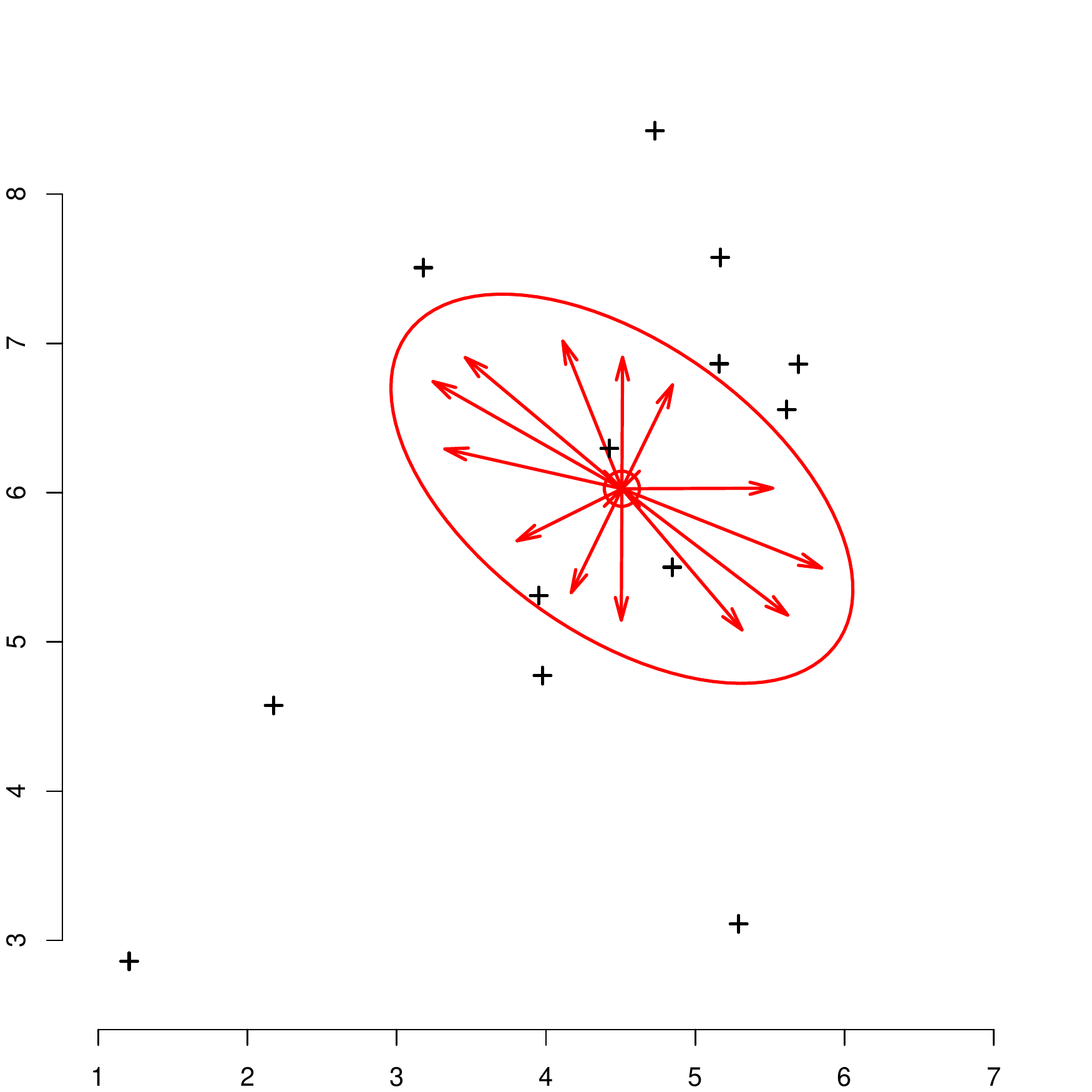}
\end{minipage}
\caption{
		Data sample together with
      {\bf a)} mean and sample covariance matrix,  {\bf b)} marginal median, marginal signs and MSCM,
         {\bf c)} spatial mean, spatial signs and SSCM and {\bf d)} Oja median, Oja signs and OSCM. }
\label{Fig:mult.signs}
\end{figure}

Multivariate signs can be useful for obtaining information about the spread of the data and dependencies among the variables. We call
\[
	\OSCM(\X; \bm) =  \frac{1}{n} \sum_{i=1}^n \osgnx(\bx_i; \bm) \osgnx(\bx_i; \bm)^\top
\]
the \emph{Oja sign covariance matrix of $\X$ w.r.t.\ the central location $\bm \in \mathbb{R}^k$},
\[
	\OSCM(\X) = \OSCM(\X; \omed(\X))  =  \frac{1}{n} \sum_{i=1}^n \osgnx(\bx_i)  \, \osgnx(\bx_i)^\top
\]
the \emph{Oja sign covariance matrix of $\X$} and
\[
	\ORCM(\X) =  \frac{1}{n} \sum_{i=1}^n \ornkx(\bx_i) \, \ornkx(\bx_i)^\top
\]
the \emph{Oja rank covariance matrix of $\X$}.
In an analogous way we define marginal and spatial sign and rank matrices $\MSCM, \SSCM$ and $\MRCM, \SRCM$ by replacing $\osgnx$ by $\msgnx$, $\ssgnx$ and $\ornkx$ by $\mrnkx$ and $\srnk$, respectively.
A comparative study of all sign and rank matrices presented here can be found in \citet{Visuri2000}. Figure \ref{Fig:mult.signs} shows a small two-dimensional data sample, together with the different multivariate medians and the corresponding signs and sign covariance matrices visualized as ellipses. The ellipses have radius $\sqrt{\chi^2_{2,0.8}}$, i.e.\ the positive definite matrix $S$ is depicted by the ellipse $\x^\top S^{-1} \x = \chi^2_{2,0.8}$.
The affine equivariance (\ref{sign.affine}) for Oja signs and ranks translates into a
similar equivariance for the corresponding covariance matrices,
\[	
	\OSCM(\X \bo A^\top + \bEins_n \bb^\top) = \det(\bo A)^2 (\bo A^{-1})^\top \OSCM(\X) \bo A^{-1}
\]
and
\[	
	\ORCM(\X \bo A^\top + \bEins_n \bb^\top) = \det(\bo A)^2 (\bo A^{-1})^\top \ORCM(\X) \bo A^{-1},
\]
for $\bb \in \mathbb{R}^k$ and $\bo A \in \mathbb{R}^{k \times k}$ with full rank. This has,
roughly speaking, the consequence that $\OSCM$ and $\ORCM$ estimate the inverse of the covariance
matrix up to scale. More specifically, \citet{Ollila2003} show that $\OSCM(\X; \bm)$ converges
to a multiple of the inverse of the covariance matrix at the $\sqrt{n}$ rate, if
$\bm$ is a $\sqrt{n}$-convergent location estimator and the data stem
from a linear transformation of a reflection and permutation invariant distribution having finite
second moments. The statistical theory of the Oja rank covariance matrix $\ORCM$ is treated in \citet{Visuri2003} and
\citet{Ollila2004}.

A further consequence of this remarkable property is that Oja sign and rank matrices provide
easily-obtained, positive definite, consistent estimates for the covariance matrix
up to scale, which is a significant advantage over the other sign and rank covariance matrices.
Multivariate data analysis is primarily aimed at analyzing the dependencies and interactions
between variables. For multivariate methods such as correlation, canonical correlation analysis,
principal component analysis, or factor analysis, it is fully sufficient
to know the covariance matrix only up to scale. In particular, the $\OSCM$ and the $\ORCM$ directly
 estimate a multiple of the \emph{concentration matrix} or \emph{precision matrix}, i.e., the inverse covariance matrix, which plays an important role in graphical models. The
application of $\OSCM$ in this context is examined in \citet{Vogel2008} and \citet{Vogel2008b}.
Due to their (inverse proportional) affine equivariance, the theory developed in \citet{Vogel2011} also applies to the $\OSCM$ and the $\ORCM$.

Furthermore \citet{Ollila2003} and \citet{Ollila2004} also derive the limiting distributions
of $\OSCM$ and $\ORCM$ in the elliptical model. Contrary to the other non-parametric covariance
matrices based on marginal and spatials signs and ranks, the OSCM and the ORCM are not
invariant w.r.t.\ the elliptical generator within the elliptical model. Their asymptotic distribution depends on the tail
behavior of the population distribution. But, and this is also in contrast to marginal and
spatial non-parametrics, Oja sign and rank matrices are very efficient at the normal model in low dimensions.
Their performance almost equals that of the empirical covariance matrix, the
normal maximum likelihood estimator. They outperform the empirical covariance matrix ECM 
at heavier
tailed distributions, but are generally less efficient at light tails. They maintain the
good efficiency at small sample sizes $n$ and small dimensions $k$, which is not true for
many robust scatter estimators.

\begin{figure}[ht]
\centering
\includegraphics[width=0.8\linewidth]{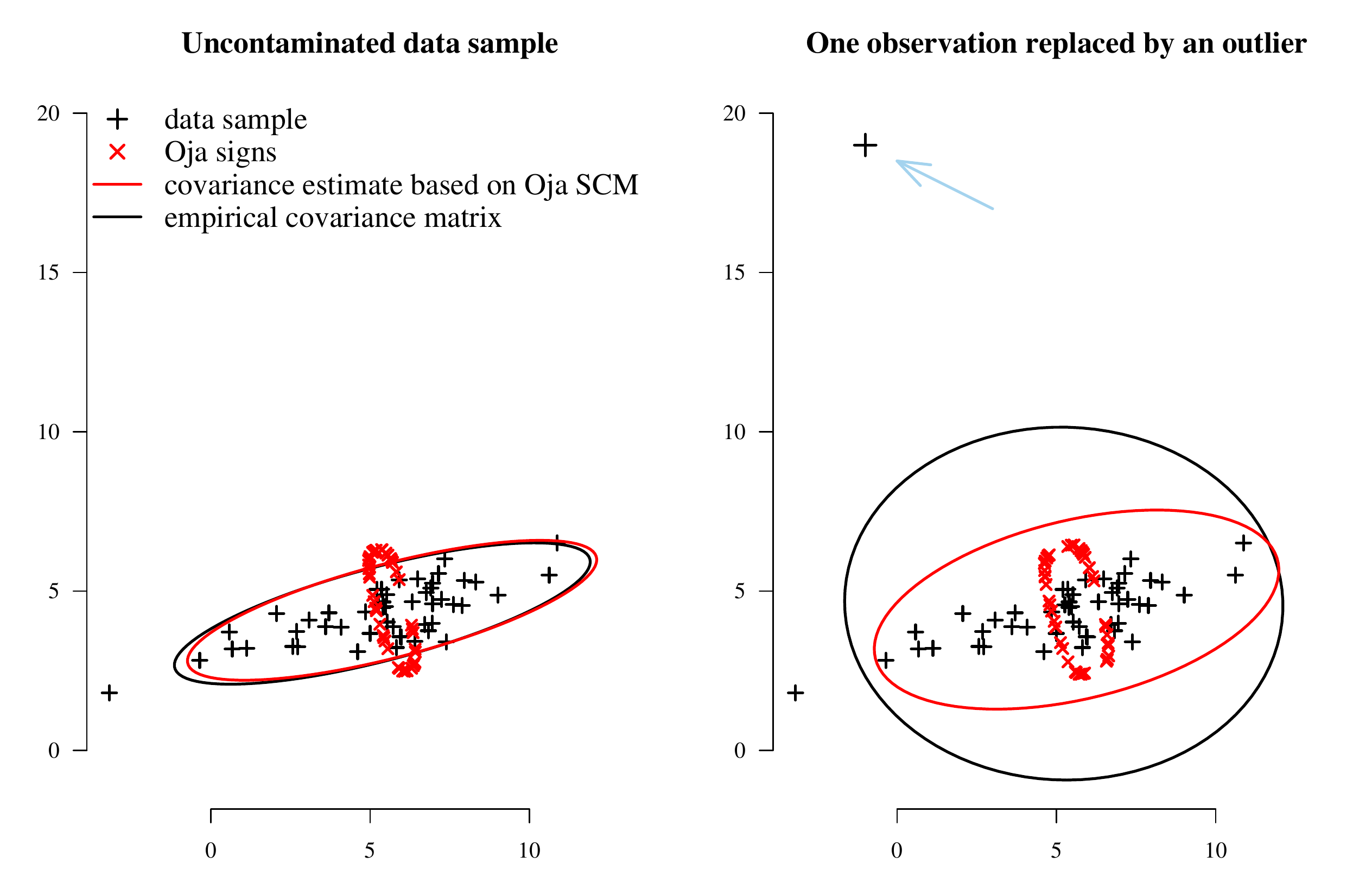}
\caption{
		The effect of an outlier on the ECM and on the OSCM-based covariance estimate. }
\label{Fig:bias.oscm}
\end{figure}

The gain in using the OSCM or the ORCM instead of the sample covariance matrix lies in their higher robustness, which comes at practically no loss in efficiency (but unfortunately at a large increase in computing time). Similar to the Oja median,
the Oja sign and rank matrices do not qualify as globally robust estimators in the sense
of having a breakdown point near 1/2. They require first moments; their influence functions
are unbounded but linear instead of quadratic (as the influence function of the sample covariance matrix), and
the asymptotic breakdown point is zero. Very few misplaced observations suffice to let
the bias of the estimators become arbitrarily large, but the bias is significantly smaller
than that of the sample covariance matrix, cf.\ Figure \ref{Fig:bias.oscm}.

\subsection[{The one- and C-sample location tests based on Oja signs and ranks}]{The one- and $C$-sample location tests based on Oja signs and ranks}
\label{subsec:1.5}

The notions of multivariate location and spread together with the corresponding concepts of sign and rank allow
the general derivation of multivariate inference methods. The general idea is there to view the signs, signed ranks and ranks as scores, which replace the observations in the classical multivariate procedures. In principle, robust counterparts of any multivariate method can be derived this way. We demonstrate here the multivariate one-sample and $C$-sample location tests.
For that purpose, denote for a sample point $\bo x_i$ the corresponding score $\bo s(\bo x_i; \bo m)$, where $\bo m$ is an optional location w.r.t.\  to which the score is computed.

\subsubsection{The one-sample tests}

Assume $\mathbb{X}=(\bo x_1,\ldots, \bo x_n)$ is a sample of size $n$ from a $k$-variate symmetric distribution $F$ with symmetry center $\bs \mu$.
We are interested in testing the null hypothesis $H_0: \bs \mu = \bs \mu_0$ against $H_1:  \bs \mu \neq \bs \mu_0$.

Denote $\bar{\bo s} = \frac{1}{n} \sum_{i=1}^n \bo s(\bo x_i, \bs \mu_0)$ as the average of the score values under the null hypothesis and $\bs \Sigma_{\bo s} = \frac{1}{n} \sum_{i=1}^n \bo s(\bo x_i, \bs \mu_0)\bo s(\bo x_i, \bs \mu_0)^\top$. The test statistic is then
\[
Q = n\, \bar{\bo s}^\top \bs \Sigma_{\bo s}^{-1} \bar{\bo s}.
\]
Using Oja signs or Oja signed ranks as scores, this yields a straightforward extension of Hotelling's classical one-sample $T^2$-test.
The test is invariant under affine transformations and asymptotically distribution-free and has a limiting $\chi^2_k$ distribution.
Test decisions can also be based on permutation principles by randomly changing the signs of the scores. The tests are described in detail in \citet{HettmanspergerNyblomOja1994} and \citet{HettmanspergerMottonenOja1997}.
Similar tests can also be naturally constructed using marginal or spatial signs and signed ranks. See \citet{PuriSen1971} and \citet{Oja2010} for details.

\subsubsection[The C-sample tests]{The $C$-sample tests}

Let $\mathbb{X}_c = (\bo x_{c,1},\ldots,\bo x_{c,n_c})$, $c=1,\ldots,C$, correspond to $k$-variate samples coming from $C \geq 2$ groups
having distributions $F_c$ that differ only in location parameters $\bs \mu_c$, $c=1,\ldots, C$. The null hypothesis is
$\bs \mu_1 = \ldots = \bs \mu_C$, i.e., the $C$ groups have the same location.

Denote $\mathbb{X} = (\mathbb{X}_{1},\ldots,\mathbb{X}_{C})$ as the combined sample and $n = \sum_{i=1}^C n_i$. Then
$\bar{\bo s}_c = \frac{1}{n_c} \sum_{j=1}^{n_c} \bo s(\bo x_{c,j})$, $c=1,\ldots,C$, is the average score value of group $c$ computed w.r.t.\  to the location of the combined sample. Similarly, $\bs \Sigma_{\bo s} = \frac{1}{n} \sum_{i=1}^n \bo s(\bo x_i)\bo s(\bo x_i)^\top$ is computed for the combined groups.

The test statistic is obtained as
\[
Q=\sum_{c=1}^C n_c \bar{\bo s}_c^\top \bs \Sigma_{\bo s}^{-1} \bar{\bo s}_c.
\]
When Oja signs and Oja ranks are used as scores, $C$-sample tests for multivariate location are obtained that are asymptotically distribution-free and affine invariant. The limiting distribution of the test statistic is $\chi^2_{k(C-1)}$, but
$p$-values can be obtained by permuting observations between the groups.
The two tests are described in \citet{HettmanspergerOja1994,HettmanspergerMottonenOja1998}. Similar tests based on other concepts of signs and ranks are described also in \citet{PuriSen1971} and \citet{Oja2010}.

\section[Algorithms]{Description of the algorithms} \label{sec:DesAlg}
The package \pkg{OjaNP} contains four different algorithms to calculate the
Oja median. Two exact algorithms and two approximate algorithms. The first exact
algorithm was developed in \citet{Ronkainen2003} as well as one of the approximate
algorithms. The second exact algorithm \citep{MoslerPokotylo2015} is based on the first one:
it accelerates the computation considerably by introducing bounds to the region of search.
The numerical calculation is a non-trivial problem which consumes enormous
calculation resources and hence, the exact algorithms are limited to small data
situations only and, as a consequence, approximate algorithms are needed.
These offer parameters to regulate the speed vs.\ accuracy trade-off, and the user has to decide from case to case
which algorithm to choose with which tuning parameters.
In Section~\ref{sec:Use}
we will give an overview over the several options and their effect onto the
calculation precision and time.
Before that, we are going to describe the four
algorithms.

\subsection{Exact algorithm}
\citet{Ronkainen2003} implemented the ideas from \citet{Niinimaa1992} and
generalized them into higher dimensions with the help of the result described
in \citet{Hettmansperger1999}, whereby the vertices of the Oja median set are always located
on intersections of hyperplanes that are spanned by data points.

\citet{Ronkainen2003} constructed a Las Vegas algorithm as follows. (This is a simplified version; a more detailed
description can be found in the original paper.)
\begin{enumerate}
 \item \verb+Let+ $\mathcal{H}$ \verb+ be the set of all+ $(k-1)$\verb+-dimensional hyperplanes spanned by the+ \\
\verb+points in+ $\mathbb{X}$.
 \item \verb+Take the data point+ $\mathbf{x}_{c} \in \mathbb{X}$ \verb+closest to the mean as an initial+\\
       \verb+candidate point+.
 \item \verb+Sample+ $k-1$ \verb+hyperplanes out of+ $\mathcal{H}$ \verb+such that the candidate point+\\
       \verb+is on their intersection+ $L$.
 \item \verb+Calculate the Oja depth of each intersection point between+ $L$ \verb+and+\\
       \verb+the hyperplanes in +$\mathcal{H}$.
 \item \verb+Take the point+ $\mathbf{x}^*_{c}$ \verb+with the highest Oja depth as next candidate point for+\\
       \verb+the Oja median+.
 \item \verb+Repeat steps 3 to 5 until no improvement in the objective function is+\\
       \verb+possible (or latest after+ $n$ \verb+repetitions)+.
 \item \verb+The result for the exact Oja median is the last candidate point+ $\mathbf{x}^*_{c}$.
\end{enumerate}

\citet{Ronkainen2003} focused on computational stability rather than efficiency.
In case a candidate point is a data point, there are $k-1$ possible
intersection hyperplanes $L$.  Instead of only following the one determined by the gradient of the objective function, the algorithm tries all possible ones. 

This algorithm finds just one of the vertices of the median set. While searching for the median, the algorithm may pass through several vertices of the median set, although it is not guaranteed that it visits all of them. The reason is that on step 5 only the first of possibly two points having highest Oja depth is taken as $\mathbf{x}^*_{c}$. However, in case of a non-unique median, there exist two such points lying on an edge of the median set. To deliver all vertices of the median set, the algorithm can be modified as follows: it has to store both points as vertices and, in addition, check all lines passing through them.

\subsection{Exact bounded algorithm}
\label{sec:eba}

Based on the exact algorithm of \citet{Ronkainen2003}, \citet{MoslerPokotylo2015} developed a faster exact algorithm.
This algorithm uses the centered rank functions to build bounded regions which contain the median.
The negative rank function $-\ornkx(\x)$ is a vector that points in a direction of 
ascent of the depth function. It defines a hyperplane through $\bo{x}$, on the positive side of which the Oja median is found.
The halfspaces defined by the negative rank function are used to build a bounded region that contains the median.
In this algorithm, these halfspaces are selected in an iterative way and the further search is restricted to their intersection.
The hyperplanes bordering such a search region will be called \textit{bounding hyperplanes} or simply \textit{bounds}.

The steps of the algorithm are as follows. (A more detailed
description can be found in the original paper):

{\ttfamily\hyphenchar\font=45\relax

\begin{enumerate}
 \item Let $\mathcal{H}$ be the set of all hyperplanes spanned by the points in $\mathbb{X}$.
 \item Create the initial rectangular bounded region $\bo{B}$,  limited by hyperplanes that are perpendicular to the coordinate axes and go through the maximal and minimal coordinates of the data points on these axes.
 \item Iteratively reduce the bounded region $\bo{B}$  by adding hyperplanes that go through a properly chosen central point of the region and have their normal vectors equal to the corresponding negative rank function. Specifically, the mean value of the bounds' intersection points is selected as a central point in our implementation. The bounds of $\bo{B}$  that are cut off by newly added hyperplanes are removed.
 \item The region is reduced until the desired final volume of the bounded region is reached.
 \item Add the bounds from $\bo{B}$ to  $\mathcal{H}$.
\item \begin{itemize}
 \item At the first iteration: Take a random initial line $L$ on the border  of $\bo{B}$.
 \item At further iterations: Sample $k-1$ hyperplanes out of $\mathcal{H}$ in that way, that the candidate point
       is on their intersection line $L$.
       \end{itemize}
 \item Calculate the Oja depth on each intersection point between $L$ and  hyperplanes in $\mathcal{H}$ that lies in the bounded region.
 \item Take the point $\mathbf{x}^*_{c}$ with the highest Oja depth as next candidate point for the Oja median.
 \item Repeat steps 6 to 8 until no improvement in the objective function is
       possible (or latest after $n$ repetitions).
 \item The result for the exact Oja median is the last candidate point $\mathbf{x}^*_{c}$.
\end{enumerate}
}

The bounded regions reduce the complexity of the searching procedure by reducing the number of hyperplanes that cross the searching lines as well as the number of their intersections to be considered in the minimization procedure.

\begin{figure}\center{
\includegraphics[scale=.4]{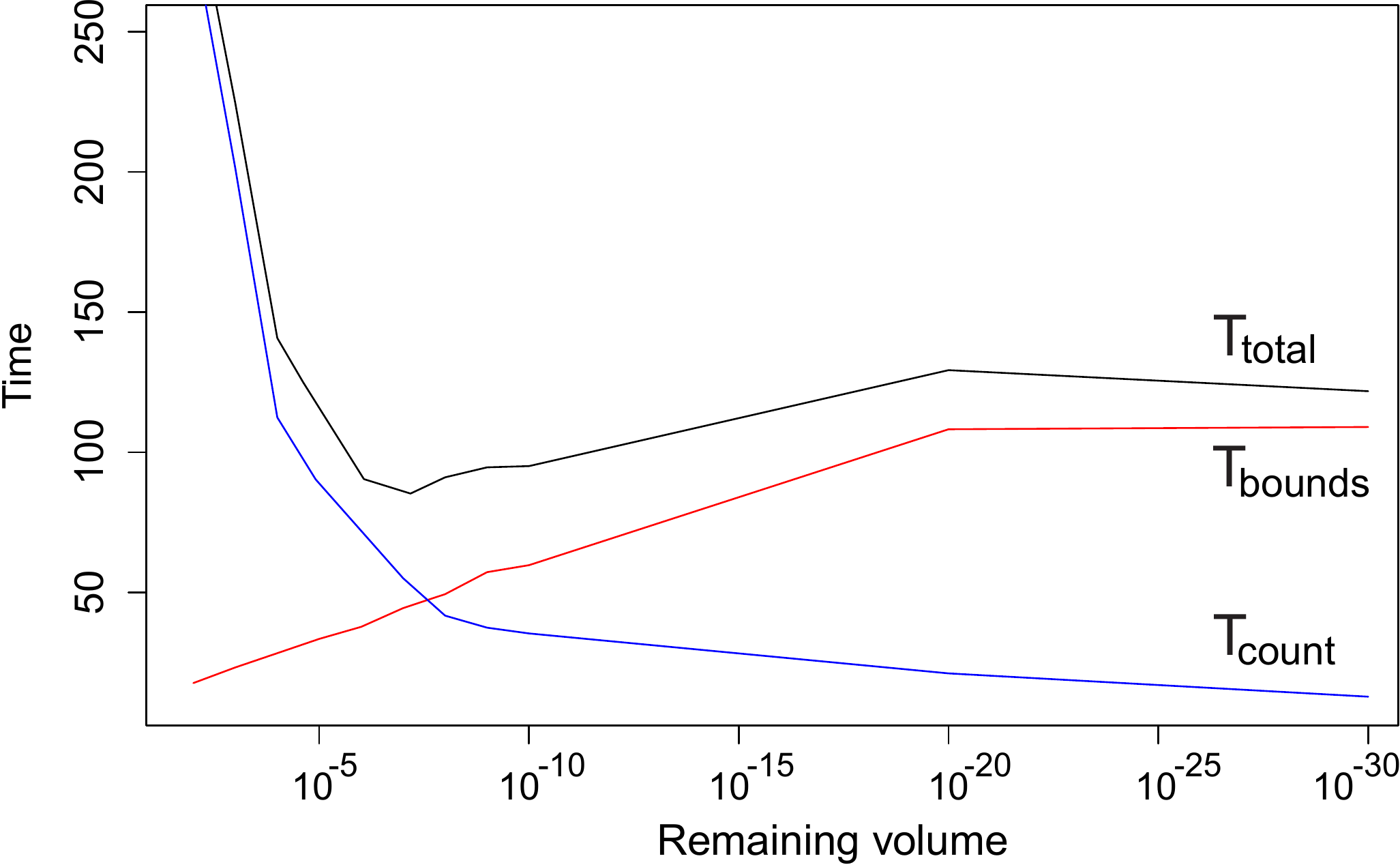}
\caption{Dependence of calculation time on the size of the bounded region; time needed for bounding ($T_{bounds}$), for minimizing ($T_{count}$), and total time ($T_{total}$). For comparison: total time of the first exact algorithm is about 340 seconds.}
\label{fig:volume_time}
}\end{figure}

The algorithm is driven by the desired final volume of the bounded region, which is the volume of the minimal rectangle containing the region and having edges parallel to the coordinate axes.
As this parameter is reduced, the time needed to build the bounded regions ($T_{bounds}$) increases, while the minimization time ($T_{count}$) decreases along with the number of hyperplanes and their intersections; see Fig.~\ref{fig:volume_time}.
The total computing time ($T_{total}$) decreases rapidly with the volume, but then slowly grows again.
 Beyond some point the procedure becomes less efficient. For comparison, the original algorithm needs a total time $T_{total}$ of ca.\ 340 seconds in this example.
 It appears that the fastest computation is obtained if bounds are imposed until the volume of the bounded region ranges around $10^{-8}$ of the original volume.
Note that the bounds may cut off some of the vertices of the median set.
Moreover, if the central point of the bounded region lies in the median set on step 3, its negative rank function is zero, and this point is directly returned as a median, as on Fig. \ref{fig:exBehavSimple}.

Due to limitations in computing memory and long calculation time, the exact algorithm and its bounded version are only able to calculate the Oja median for small datasets in low dimensions. For example, the calculation of the median in a data set of size $100 \times 5$ needs 12 GB RAM.
Therefore, approximate algorithms are needed.

Obviously, the bounded algorithm can be stopped at any iteration and some mean value of the last bounded region be taken as an approximation of the Oja median. However, unlike the approximate algorithms presented below, this approach requires the calculation of all hyperplanes. Due to its high computational requirements it is less suited as an approximate algorithm for big data sets in high dimensions.

\subsection{Grid-based algorithm}
The third algorithm, which calculates an approximation to the Oja median, was
also proposed in \citet{Ronkainen2003}. Technically it is a Monte Carlo algorithm.
The algorithm lays a uniform grid over the dataset. At each grid point a test is performed whether the point is a possible candidate.
The amount of candidate points is reduced as long as only one grid point is left. This point is afterwards the
center for a smaller but denser grid, where again each grid point is tested.
The algorithm stops when the distance between two grid points gets smaller
than a predefined parameter. A second tuning parameter is the significance level of the point tests. The steps of the algorithm are:
\begin{enumerate}
 \item \verb+Create a grid + $\mathcal{G}$\verb+ with equidistant knot distance+ $h$\verb+, covering the whole+ \\
       \verb+dataset.+
 \item \verb+Choose randomly a set of hyperplanes, build the test statistic and test+\\
       \verb+each of the grid knots in + $\mathcal{G}$\verb+ whether it is an Oja median.+
 \item \verb+Remove those grid knots which have been tested not to be an Oja median.+
 \item \verb+If there is more than one knot left, sample additional hyperplanes and+\\
       \verb+repeat the test for the remaining knots.+
 \item \verb+Repeat these steps until only one grid knot is left over. If the last+\\
       \verb+test removes all remaining ones, take the last set.+
 \item \verb+Build a new grid around the last remaining old grid knot with+\\
       \verb+equidistant knot distance+ $h/2$\verb+.+
 \item \verb+Repeat all these steps until the grid distance reaches a predefined+\\
       \verb+threshold.+
 \item \verb+The last point is taken as the Oja median.+
\end{enumerate}

For further details, especially about the testing procedure, we refer to the original paper \citet{Ronkainen2003}.
It may happen that there is continually more than one point left on step 5. In this case sampling of the additional hyperplanes may not help and the algorithm hangs.
We restrict the algorithm to 5000 iterations on step 5, after which the grid point with the best test statistic
is passed to step 6 and the number of iterations is reduced to 100. We repeat until the grid threshold is reached and return the grid point with the best test statistic as an Oja median approximation.

\subsection{Evolutionary algorithm}
The fourth algorithm to calculate the Oja median is an evolutionary algorithm,
which is based on mutations of the latest candidate points. It was developed
by the Department of Computer Science, Efficient Algorithms and Complexity
Theory at the TU Dortmund, but has not been published before. The algorithm
works as follows:

\begin{enumerate}
\item \verb+Set the level of initial mutation variance+ $\sigma^2_0$.
\item \verb+Take 10 randomly chosen observations from+ $\mathbb{X}$.
\item \verb+Evaluate the objective function for all these points and take the minimum+\\
      \verb+as starting candidate point+ $\eta$.
\item \verb+Choose+ $k+1$ \verb+random numbers+ $x_1,\dots,x_k,l$ \verb+from a+ $N(0,\sigma_0^2)$\verb+ distribution and+\\
      \verb+calculate the+ $k$\verb+-variate mutation vector+\\
      $$\mathbf{\nu}=\frac{|l|}{\sqrt{x_1^2 + \dots + x_k^2}}(x_1,\dots,x_k)^\top\,.$$
      \verb+The mutation vector+ $\mathbf{\nu}$ \verb+has a normally distributed length with given variance+\\
      $\sigma^2_0$ \verb+and uniformly distributed direction+.
\item \verb+Calculate+ $m$ \verb+mutation points+ $\mathbf{\eta}_i=\mathbf{\eta} + \mathbf{\nu}_i$ \verb+for+ $i=1,\dots,m$ \verb+from the last+\\
      \verb+candidate point+ $\mathbf{\eta}$.
\item \verb+Calculate the ratio+ $r$ \verb+how often the objective function is bigger+\\
      \verb+at the mutations than at+ $\mathbf{\eta}$.
\item \verb+If+ $r>0.2$ \verb+then+ $\sigma_0^2 * \kappa$\verb+ else+ $\sigma_0^2 * \frac{1}{\kappa}$ \verb+for some+ $\kappa > 1$\verb+.+
\item \verb+Choose as a new candidate point the mutation with the smallest objective+\\
      \verb+function value+.
\item \verb+Repeat steps 4 to 8 until the variance for the next mutation drops+\\
      \verb+under a predefined value+ $s$.
\item \verb+If the algorithm has not terminated after+ $n_t$ \verb+steps, stop the calculation.+
\end{enumerate}
Step 7 controls the dynamic of the mutation. If at more than
20\% of the mutations the objective function has a smaller value than at the last candidate point, the algorithm
increases the variability of the mutation; hence the search area is enlarged. Step 10 ensures that the algorithm terminates in any case.

 \subsection[Other algorithms for the Oja median in R]{Other algorithms for the Oja median in \proglang{R}}
 
 There are other implementations of algorithms for the Oja Median
 available in \proglang{R}, but they are mostly restricted to two dimensions.
 
  The function \code{med} in the package \pkg{depth} \citep{DepthPackage}
 uses the \proglang{Fortran} code of \citet{Niinimaa1992} and is restricted to
 the bivariate case.

 Another method to compute the Oja Median was suggested by Roger Koenker on
 \proglang{R}-help on 16. Aug 2003 (see \url{http://tolstoy.newcastle.edu.au/R/help/03b/1990.html}
 using the \pkg{quantreg} package \citep{quantregPackage}:
 
  \begin{verbatim}
  oja.median <-function(x)
  {
    ##
    ## bivariate version -- x is assumed to be an n by 2 matrix
    ##
    require(quantreg)
    n <- dim(x)[1]
    A <- matrix(rep(1:n, n), n)
    i <- A[col(A) < row(A)]
    j <- A[n + 1. - col(A) > row(A)]
    xx <- cbind(x[i,  ], x[j,  ])
    y <- xx[, 1] * xx[, 4] - xx[, 2] * xx[, 3]
    z1 <- (xx[, 4] - xx[, 2])
    z2 <-  - (xx[, 3] - xx[, 1])
    return(rq(y~cbind(z1, z2)-1)$coef)
  }
  \end{verbatim}

\section[OjaNP package]{The R package \pkg{OjaNP}} \label{sec:Use}

\begin{figure}[ht]
\centering
\includegraphics[width=0.7\textwidth]{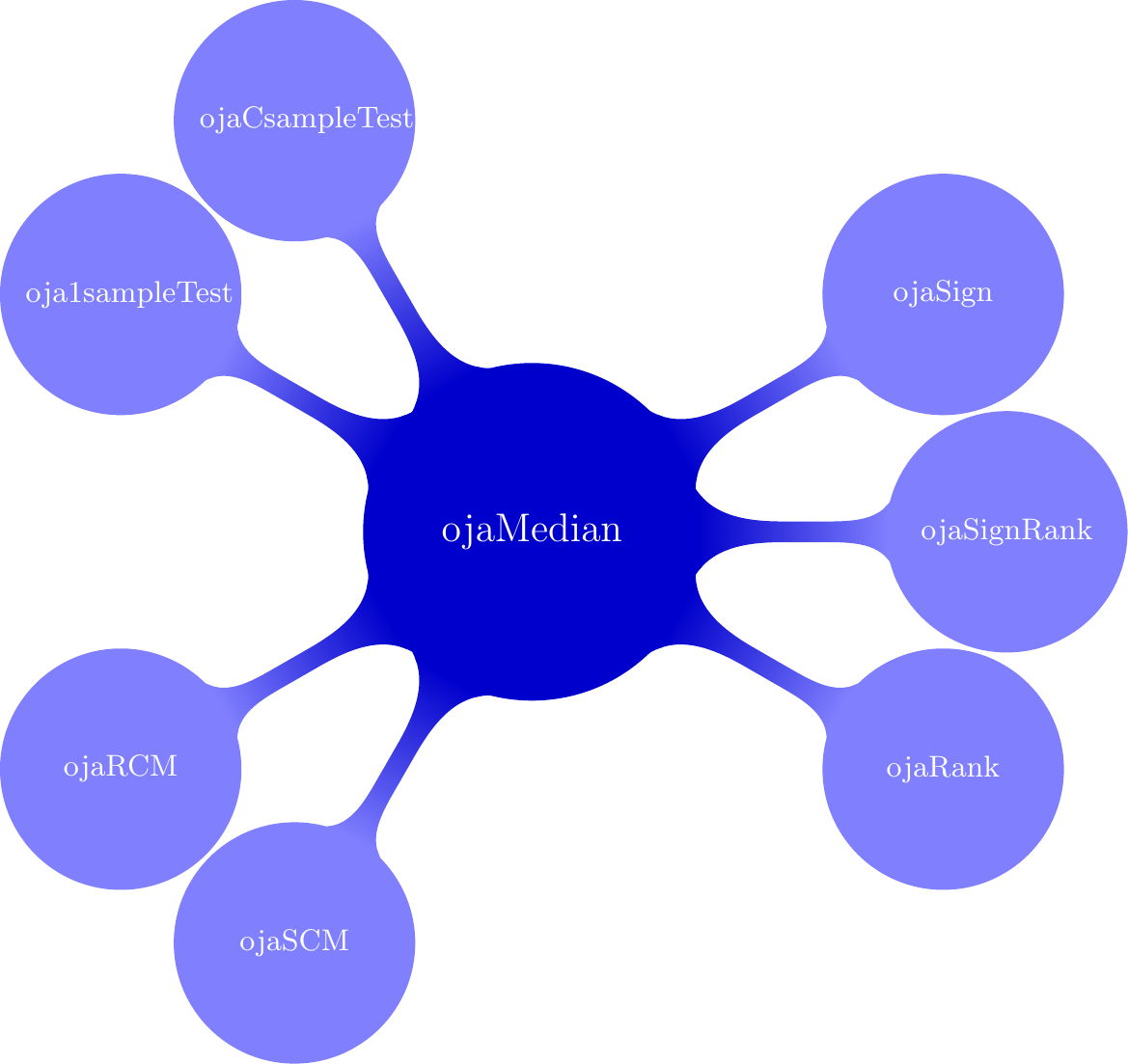}
\caption{The main functions in the package \pkg{OjaNP}.}
\label{fig:Schema}
\end{figure}

The main purpose of the \pkg{OjaNP} package is to provide users with the possibility to compute the Oja median. The package includes, however, also other useful functions. The main functions of the package are visualized in Figure~\ref{fig:Schema}.

Most of the function names are self-explanatory. For details about them we refer to the corresponding help pages.
In the following we will first explain the function \code{ojaMedian} and its options in detail. Then we demonstrate the use of some of the functions with a small but illustrative data set, which is also contained in the package.

\subsection[ojaMedian]{The computation of the Oja median in \pkg{OjaNP}} \label{sec:Use2}

The main function of \pkg{OjaNP} is
\begin{center}
\code{ojaMedian(X, alg = "evolutionary", sp = 1, na.action = na.fail,\\
        \quad control = ojaMedianControl(...), ...)}
\end{center}

The user can choose via the \code{alg} option between four algorithms to calculate the
Oja median. Furthermore, we have an option to calculate the Oja median repeatedly and
average these results in order to receive less varying results. The amount of repetitions
can be controlled with the \code{sp} parameter.

In what follows we are going to explain the different parameters which control
the flow of the different algorithms in detail and give insights how to choose the parameters in
a given data situation. The default algorithm of the \code{ojaMedian} function
is the evolutionary algorithm.

The evolutionary algorithm loses the affine equivariance property of the Oja median.
In order to restore it we first perform a scatter matrix transformation to obtain an
invariant coordinate system (implemented in \pkg{ICS} \citet{IcsPackage}), apply the algorithm to the
transformed data and re-transform afterwards. That way we restore the affine equivariance
for this implementation.

\begin{figure}[ht]
\centering
\includegraphics[width=0.8\textwidth]{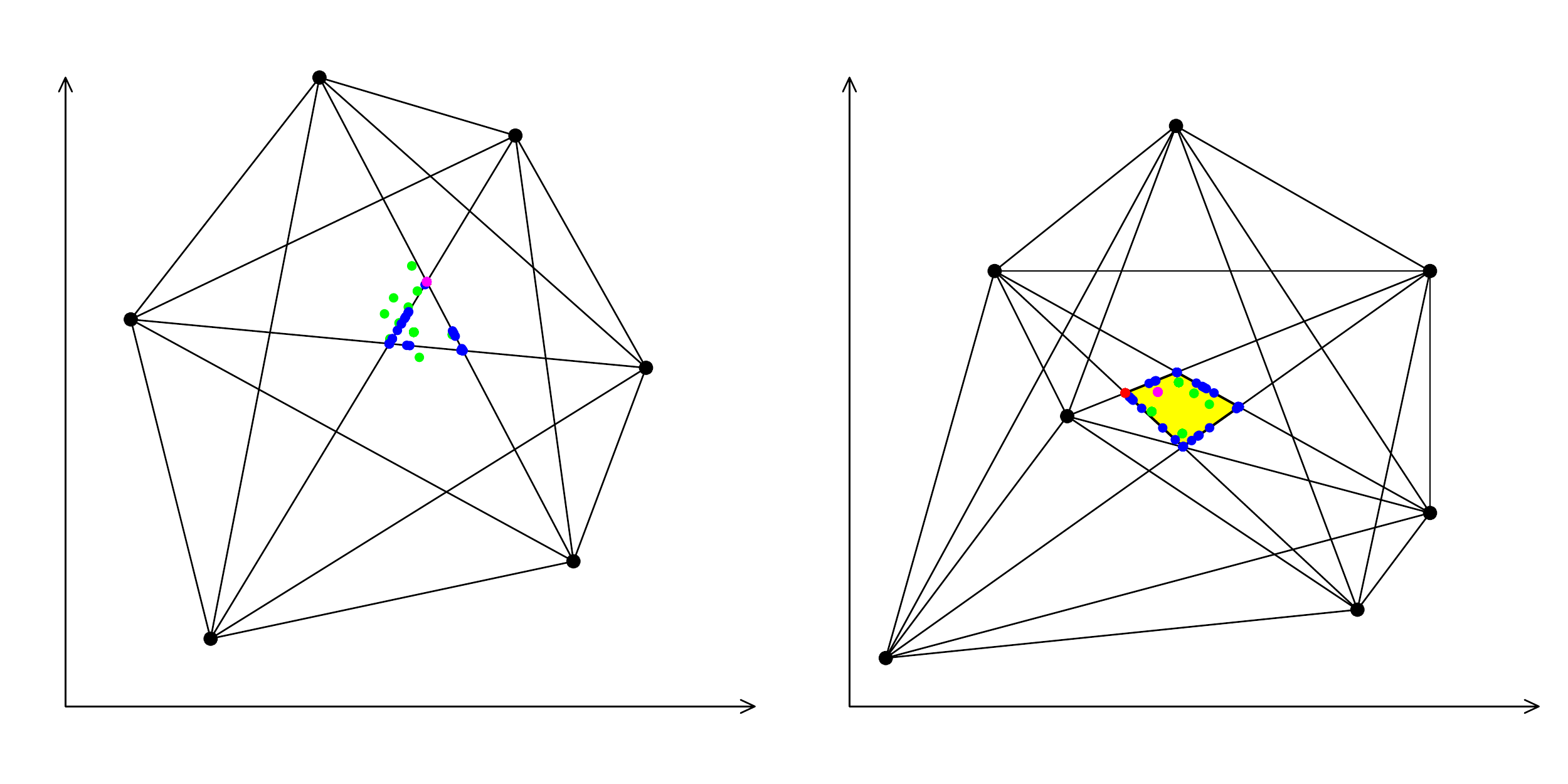}
\caption{Example plot for all four algorithms: exact algorithms (red), evolutionary algorithm (blue), grid algorithm
(green). The convex median set is marked with yellow.}
\label{fig:exBehavSimple}
\end{figure}

Figure \ref{fig:exBehavSimple} shows the exemplary outcome of the four implemented
algorithms in simple data situations for the unique (left) and non-unique (right) case. We have chosen simple
data situations with 6 and 7 data points, run the different algorithms 500 times and
plotted the outcome into the figure.

In the left part of Figure \ref{fig:exBehavSimple} we have a data set that has a unique
Oja median. This one is correctly determined by the exact implementations (red), whereas both
approximate algorithms have a systematic behavior which does not differ strongly from
the non-unique case in the right side of the graphic. The evolutionary algorithm (blue)
determines the Oja median always along lines of intersection with the result of a bordered
area.
The right-hand side of Figure \ref{fig:exBehavSimple} exhibits data that have a non-unique Oja median.
Here the two exact algorithms find a vertex (red) of the median set, while the evolutionary algorithm (blue) yields any point of
the border of the median set, that is the area with the lowest Oja depth (yellow). The grid algorithm, however terminates in this 
case usually within the convex median set.

In a next step we are going to analyze the outcome of the algorithms in more complex data
situations. The first typical data situation is a multivariate normal distributed data cloud
and we calculate the Oja median with the exact (red), the grid (green) and the evolutionary
algorithm (blue). As we can see the evolutionary algorithm has the biggest variation within
the results, but we have to keep in mind that the default setting of our function is tuned
to calculate fast results. In the next paragraph we will discuss how to get more precise
results in trade-off for calculation time. For most applications the faster, less accurate settings appear to be preferable.

The second data set of interest consists of two data clouds, which are far away from each
other. The motivation for that is to see whether the algorithms take this into account or if they get
stuck in one cloud. As you can see in the right part of the figure, that all included
algorithms are capable of calculating the Oja median to be in the center between both data
clouds.

\begin{figure}[ht]
\centering
\includegraphics[width=0.8\textwidth]{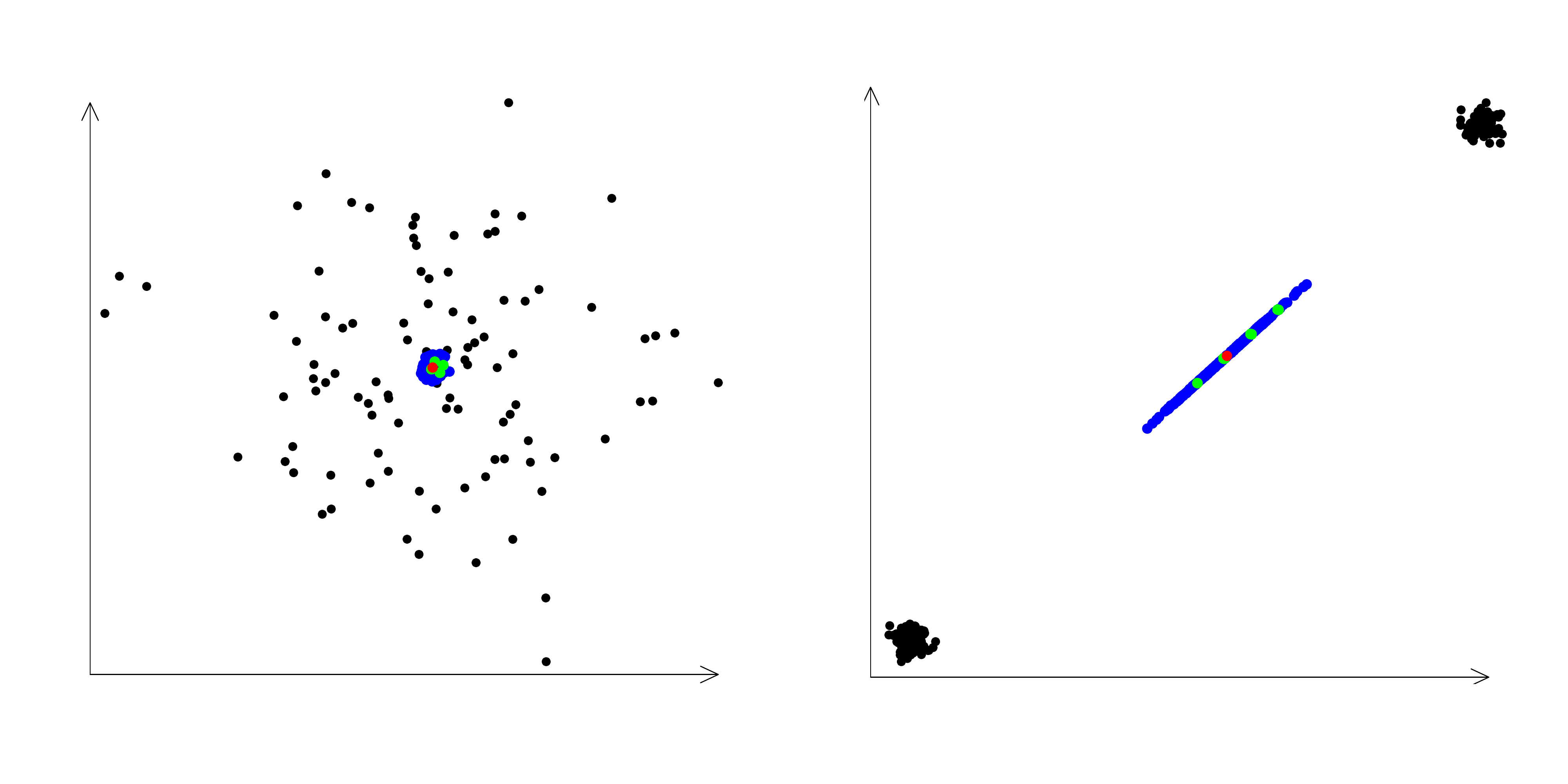}
\caption{Example plot for all four algorithms: exact algorithms (red), evolutionary algorithm (blue), grid algorithm
(green).}
\label{fig:exBehavComplex}
\end{figure}

Let us now compare the runtime of the two approximate algorithms depending on the dimension
and size of the data in the bivariate case.
We do not perform a runtime analysis for the exact algorithms. Their performance depends strongly on the computer used (particularly memory), much more so than the approximate ones. 

On an average computer, bivariate problems up to 1000 observations are solvable in acceptable time,
but for higher dimensions and sizes this decision has to be made from case to case.
For example in the seven-dimensional case with tens of observations it takes minutes for the
exact algorithm to find the solution. The biggest limitation of the exact algorithm
is the memory to store all hyperplanes. Even if we would allow infinite calculation time, the algorithm
would still not be able to calculate the exact Oja median in more complex data situations because
it cannot pre-calculate and access the total amount of hyperplanes, and hence we are facing
a corresponding address space problem. This also applies to the exact bounded algorithm.
Compared with the original exact algorithm, the bounded one finds the solution approximately two to five times faster.

\begin{figure}[ht]
\centering
\includegraphics[width=0.8\textwidth]{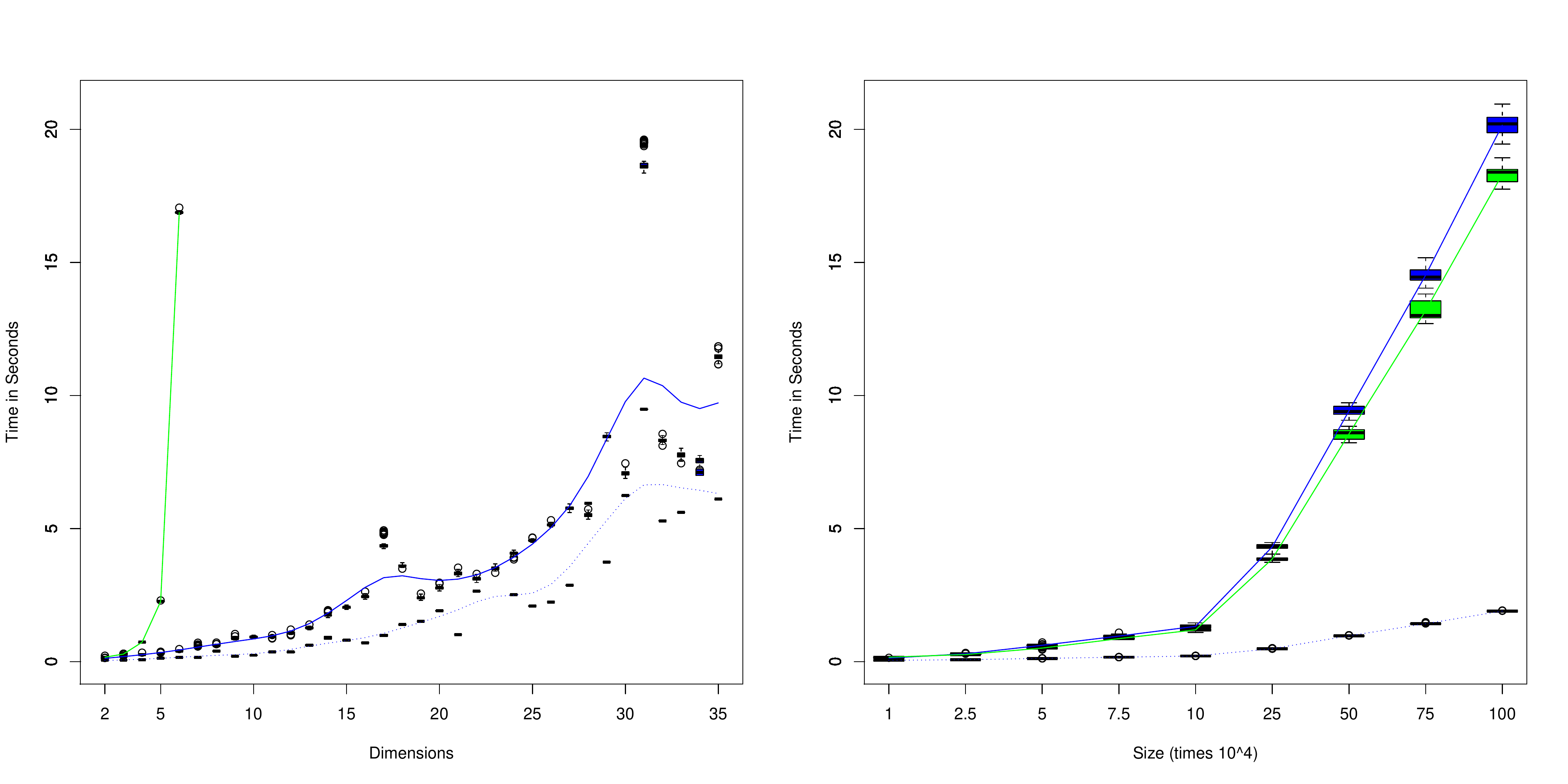}
\caption{Runtimes for the approximate algorithms: the approximate grid algorithm (solid green line), the evolutionary algorithm (solid blue line), the raw method without transformations or validation checks (dotted blue line).}
\label{fig:runtimeApprox}
\end{figure}

In order to analyze the runtime for different dimensions we simulated 10,000 multivariate
normal distributed random numbers (with $\mathbf{\mu}=\mathbf{0}$,$\mathbf{\Sigma}=\mathbf{I}$).
The approximate grid algorithm (solid green line) is only able to calculate the Oja median up
to 5 dimensions in acceptable time for this amount of data; this is why we did not take higher
dimension situations into consideration. The evolutionary algorithm (solid blue) can calculate
in the same time the Oja median of a 35 dimensional data set, and even higher dimensional
problems are solvable. Since the ICS step (and there especially the data validation checks)
takes a lot of computational time we implemented also a raw command to access the algorithms
directly without transformations or validation checks (dotted blue line). In the analysis of
different dimensions the raw algorithm does not bring huge advantages, but in the runtime
analysis concerning the size of the data in the bivariate case (right-hand side of Figure
\ref{fig:runtimeApprox}) we can detect a huge advantage for more than $10^5$ observation.
The raw algorithm is even able to calculate the Oja median for sample size $5\cdot 10^7$.
Hence, our advice in time-critical situations with high sample size is to use the raw method.
If the affine equivariant property is still required, we advise to perform beforehand ICS
separately without performing the included data validation steps.


The evolutionary algorithm has many tuning parameters, some of which control its
accuracy. As we have seen in Figure~\ref{fig:exBehavComplex}, the default settings
for these tuning parameters are pre-set to deliver fast results. In trade-off for higher
computational time, the user can adjust the settings to a more precise algorithm.
The key
parameters \emph{useAllSubsets}, \emph{nSubsetsUsed}, and \emph{sigmaLog10Dec}.
The latter is the main abort criterion of the algorithm.
It forces the algorithm to stop if the logarithmized initial variance differs more then
the value of \emph{sigmaLog10Dec} from the actual logarithmized variance. In other words,
when the variance of the mutation vector is getting small enough, the algorithm stops.

The settings for \emph{useAllSubsets} and \emph{nSubsetsUsed} take control over how
many spanned hyperplanes should be taken into account during the calculation of the
Oja median. Since the total amount of possible hyperplanes could be huge (it is
$\binom{n}{k}$ for $n$ observations in the $k$ variate case) the flag for
\emph{useAllSubsets} should be used carefully. It is more advisable to control this
with the \emph{nSubsetsUsed}. Raise this value together with \emph{sigmaLog10Dec}
for more precise values, lower them for faster results.

The dynamics of the evolutionary algorithm are controlled via the parameters
\emph{sigmaInit}, \emph{sigmaAda} and \emph{adaFactor}. All these take control
over the variance adjustments of the mutation vector. The parameter \emph{sigmaInit}
sets the initial variance of the the mutation, the settings for \emph{sigmaAda}
control after how many mutation steps the mutation variance is adjusted and
\emph{adaFactor} defines how the variance is adjusted. In most cases the default
settings work nicely.

To conclude this subsection, we would like to mention that there are many \proglang{R}-packages
available to compute various medians: The \pkg{depth} package \citep{DepthPackage} contains Tukey's median, Liu's median, spatial, marginal, and also the Oja median. The authors use the early \proglang{Fortran} implementation by \citet{Niinimaa1992}. Other packages containing different algorithms to compute the
spatial median are, e.g., \pkg{ICSNP} \citep{IcsnpPackage}, \pkg{MNM} \citep{MnmPackage,MnmPackage2} and \pkg{pcaPP} \citep{pcaPPpackage}. Some of these packages also offer functions for multivariate signs and ranks and methods based upon them.


\subsection[Demonstration]{Short demonstation of the package's main function} \label{sec:Demo}
In this section we will demonstrate the main functions of the package using the \code{biochem} data set.
This data set consists of the amounts of two chemical components in the brain measured at 22 mice. Ten of the mice belong to a control group
and twelve received a drug.

This is a very basic example just to demonstrate the basic use of the main functions. For details about the functions see also their corresponding help pages.

We first load the package and the data and create data objects for easier handling as well as fixing the random seed for reproducibility.
\begin{Schunk}
\begin{Sinput}
> library("OjaNP")
> data("biochem")
> set.seed(1)
> X <- as.matrix(biochem[,1:2])
> GROUP <- biochem$group
> GRlabel <- as.numeric(GROUP)
\end{Sinput}
\end{Schunk}

Next we compute the bivariate Oja median of the two components using the default evolutionary algorithm.
\begin{Schunk}
\begin{Sinput}
> OMev <- ojaMedian(X)
> OMev
\end{Sinput}
\begin{Soutput}
comp.1 comp.2
 1.150  0.425
\end{Soutput}
\end{Schunk}

As this toy data set is quite small, it is no problem to compute here also the exact Oja median using either of the algorithms provided.
\begin{Schunk}
\begin{Sinput}
> OMex <- ojaMedian(X, alg = "exact")
> OMex
\end{Sinput}
\begin{Soutput}
   comp.1    comp.2
1.1515385 0.4269231
\end{Soutput}
\end{Schunk}

\begin{Schunk}
\begin{Sinput}
> OMbo <- ojaMedian(X, alg = "bounded_exact")
> OMbo
\end{Sinput}
\begin{Soutput}
   comp.1    comp.2
1.1515385 0.4269231
\end{Soutput}
\end{Schunk}

\begin{figure}
\begin{center}
\includegraphics{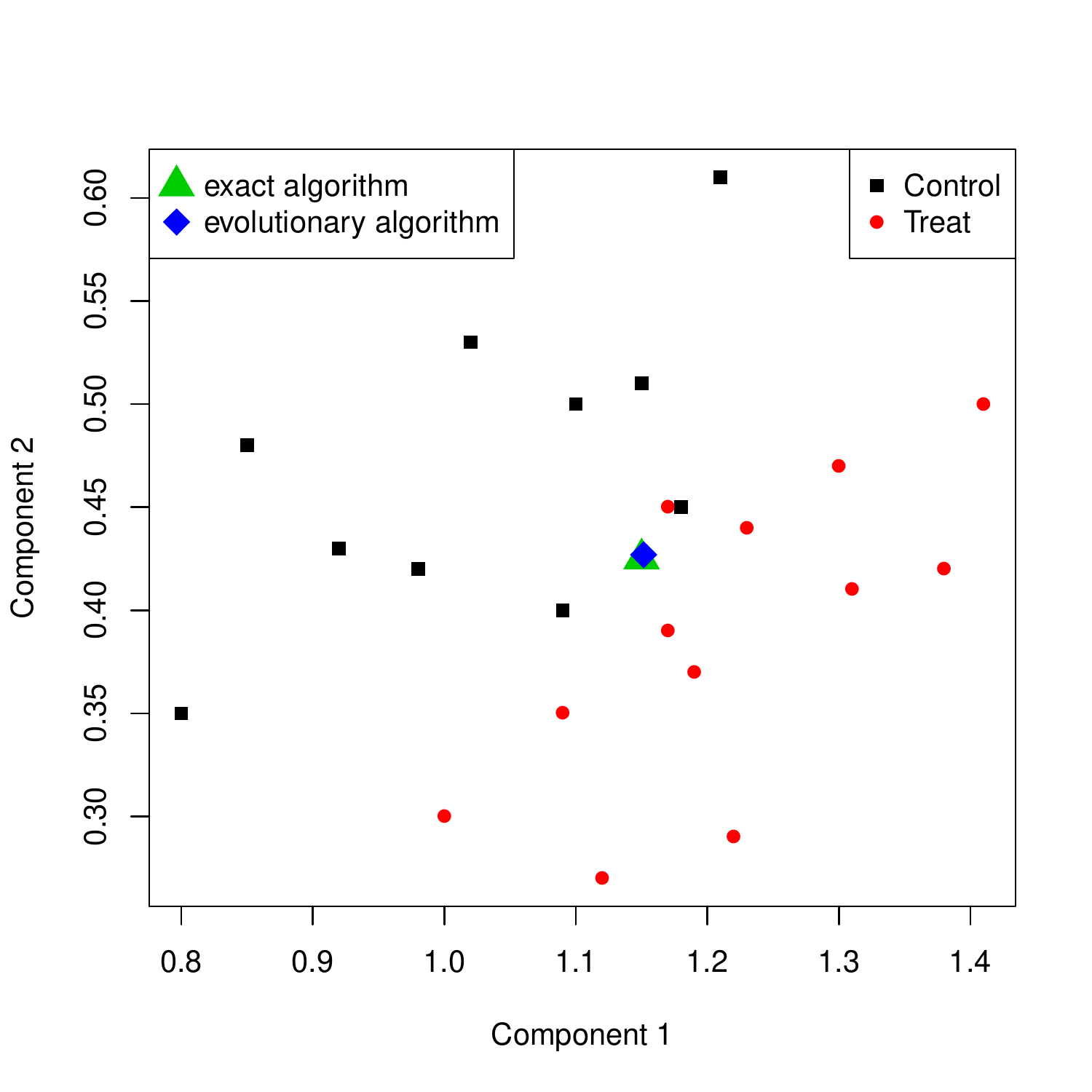}
\end{center}
\caption{Visualization of the \code{biochem data} and of the Oja median when computed using the exact algorithm and the evolutionary algorithm.} \label{OplotMS}
\end{figure}

As can be seen, the difference between the exact and the approximate estimate is rather small, which is also visualized in
Figure \ref{OplotMS}, produced by the following code:

\begin{Schunk}
\begin{Sinput}
> plot(X[,1], X[,2], col = GRlabel, pch = GRlabel + 14,
+   xlab="Component 1", ylab="Component 2")
> points(OMev[1], OMev[2], cex = 2, pch = 17, col = 3)
> points(OMex[1], OMex[2], cex = 2, pch = 18, col = 4)
> legend("topright", legend = levels(GROUP), col = 1:2, pch = 15:16)
> legend("topleft", legend = c("exact algorithm", "evolutionary algorithm"),
+   col=3:4, pch = 17:18, pt.cex = 2)
\end{Sinput}
\end{Schunk}

Next we look at the Oja signs of the data.
\begin{Schunk}
\begin{Sinput}
> head(ojaSign(X))
\end{Sinput}
\begin{Soutput}
              [,1]          [,2]
 [1,]  0.010681818  0.1150000000
 [2,] -0.063181818  0.0295454545
 [3,] -0.058409091 -0.0300000000
 [4,] -0.058863636  0.0786363636
 [5,] -0.063409091  0.0009090909
 [6,] -0.015681818  0.1204545455
\end{Soutput}
\end{Schunk}

These signs are computed with respect to the Oja median. But the \code{ojaSign} function has several options to compute them
also with respect to some other location. For example the vector of marginal medians can be specified as
\begin{Schunk}
\begin{Sinput}
> head(ojaSign(X, center = "compMedian"))
\end{Sinput}
\begin{Soutput}
              [,1]          [,2]
 [1,]  0.010681818  0.1145454545
 [2,] -0.063409091  0.0195454545
 [3,] -0.059772727 -0.0240909091
 [4,] -0.058863636  0.0781818182
 [5,] -0.063409091  0.0004545455
 [6,] -0.015681818  0.1200000000
\end{Soutput}
\end{Schunk}

The Oja signs covariance matrix can be similarly obtained as
\begin{Schunk}
\begin{Sinput}
> ojaSCM(X)
\end{Sinput}
\begin{Soutput}
              comp.1        comp.2
comp.1  0.0020624226 -0.0004861058
comp.2 -0.0004861058  0.0078892444
\end{Soutput}
\end{Schunk}

Next we test whether the Oja median of the control group corresponds to the value \code{c(1,0.5)}.
\begin{Schunk}
\begin{Sinput}
> oja1sampleTest(X[1:10, ], mu = c(1,0.5))
\end{Sinput}
\begin{Soutput}
        OJA 1 SAMPLE SIGN TEST

data:  X[1:10, ]
Q.S = 3.3745, df = 2, p-value = 0.185
alternative hypothesis: true location is not equal to c(1,0.5)
\end{Soutput}
\end{Schunk}

The test decision is here based on the limiting distribution. The sample size is rather small in this example, and one may prefer to use permutation $p$-values. Using the \code{method} argument of the function, $p$-values can be computed by permutation.
\begin{Schunk}
\begin{Sinput}
> oja1sampleTest(X[1:10, ], mu = c(1,0.5), method = "permutation")
\end{Sinput}
\begin{Soutput}
        OJA 1 SAMPLE SIGN TEST

data:  X[1:10, ]
Q.S = 3.3745, replications = 1000, p-value = 0.189
alternative hypothesis: true location is not equal to c(1,0.5)
\end{Soutput}
\end{Schunk}

To demonstrate the $C$-sample location test, we use the rank test to test whether the two groups differ and want to base the decision
on permutation principles.

\begin{Schunk}
\begin{Sinput}
> ojaCsampleTest(X~GROUP, scores="rank", method = "permutation")
\end{Sinput}
\begin{Soutput}
        OJA C SAMPLE RANK TEST

data:  X by GROUP
Q.R = 15.17, permutations = 1000, p-value = 0.001
alternative hypothesis: true location difference is not equal to c(0,0)
\end{Soutput}
\end{Schunk}


\section[Conclusions]{Conclusions} \label{sec:Conc}
There are many different multivariate medians.
In this paper we explained how the different medians extend different properties of the univariate median to the multivariate case.
The Oja median has very convincing statistical properties, but
is also among the computationally more challenging ones.
We described the \proglang{R}-package \pkg{OjaNP}, which provides four different algorithms for its computation. Along with the concept of the Oja median comes the notion of Oja signs and ranks and multivariate scatter estimators based upon them. The package provides also functions for these useful tools, which can then be used for robust multivariate inferential procedures.
As examples, we described and implemented the one-sample and the $C$-sample location test based on Oja signs and ranks.

\section*{Acknowledgements}
The work of Klaus Nordhausen was supported by the Academy of Finland (grant 268703). Oleksii Pokotylo is supported by the Cologne Graduate School of Management, Economics and Social Sciences. The work of Daniel Vogel was supported by the DFG collaborate research grant SFB 823.

\bibliography{ojaNP-references}

\end{document}